\documentclass[iop,apj,numberedappendix,twocolappendix]{emulateapj}
\usepackage{mathptmx} 
\usepackage[utf8]{inputenc}
\usepackage{amsmath}
\usepackage{amsfonts}
\usepackage{amssymb}
\usepackage{natbib}
\usepackage{graphicx}
\usepackage{aas_macros}
\usepackage[usenames,dvipsnames]{color}
\usepackage{xspace}
\usepackage{bm}
\usepackage{nicefrac}

\usepackage[normalem]{ulem} %
\usepackage{enumitem} 
\setlist{listparindent=\parindent,leftmargin=*}

\usepackage{cancel}

\definecolor{webgreen}{rgb}{0,.5,0}
\definecolor{webbrown}{rgb}{.5,.0,0.17}
\definecolor{purple}{rgb}{0.5,0,.5}

\DeclareMathOperator*{\dv}{d\!}
\DeclareMathOperator*{\Dv}{D}
\newcommand{\revision}[1]{{#1}} %

\newcommand\vect[1]{\pmb{#1}}			
\newcommand{\CR}{_\mathrm{cr}}  
\newcommand{\A}{_\mathrm{A}}  
\newcommand{\gas}{_\mathrm{g}}

\newcommand{\cm}{\,{\rm cm}}    
\newcommand{\km}{\,{\rm km}}    
\newcommand{\m}{\,{\rm m}}      
\newcommand{\pc}{\,{\rm pc}}     
     
\newcommand{\kpc}{\,{\rm kpc}}

\newcommand{\g}{\,{\rm g}}

\newcommand{\msun}{\,\text{M}_\odot}

\newcommand{\s}{\,{\rm s}}      
\newcommand{\yr}{\,{\rm yr}}    
\newcommand{\Myr}{\,{\rm Myr}}  
\newcommand{\Gyr}{\,{\rm Gyr}}

\newcommand{\kms}{\km\s^{-1}}

\newcommand{\muG}{\,\mu{\rm G}}

\newcommand{\pencil}{\textsc{pencil}\xspace}

\newcommand{\rad}{\,{\rm rad}} 
\newcommand{\ecr}{\epsilon\CR}
\newcommand{\Fcrv}{{F}}

\usepackage[breaklinks,colorlinks,
  urlcolor=blue,citecolor=blue,linkcolor=blue,pdfborder={0 0 0.1}]{hyperref}
\hypersetup{
pdftitle={The Parker instability in disk galaxies},
pdfauthor={
            L.~F.~S.~Rodrigues (orcid.org/0000-0002-3860-0525),
            G.~R.~Sarson, 
            A.~Shukurov (orcid.org/0000-0001-6200-4304),
            P.~J.~Bushby, 
            A.~Fletcher, 
          }
}
\shorttitle{Parker instability in disk galaxies}
\shortauthors{L.~F.~S.~Rodrigues, G.~R.~Sarson, A.~Shukurov, P.~J.~Bushby,
A.~Fletcher}

\submitted{Accepted for publication in the Astrophysical Journal}

\begin{document}
\begin{abstract}
We examine the evolution of the Parker instability in galactic disks using
3D numerical simulations.
We consider a local Cartesian box section of a galactic disk,
where gas,
magnetic fields and cosmic rays are all initially in
a \revision{magnetohydrostatic} equilibrium.
This is done for different choices of initial cosmic ray density and magnetic field.
The growth rates and characteristic scales obtained from the models,
as well as their dependences on the density of cosmic rays and magnetic fields,
are in broad agreement with previous (linearized, ideal) analytical work.
\revision{However, this non-ideal} instability develops a multi-modal 3D structure, which cannot be quantitatively predicted from \revision{the earlier linearized studies}.
This 3D signature of the instability will be of importance in interpreting
observations.
As a preliminary step towards such interpretations,
we calculate synthetic polarized intensity and Faraday rotation measure maps,
and the associated structure functions of the latter, from our simulations;
these suggest that the correlation scales
inferred from rotation measure maps are a possible probe for the cosmic ray content
of a given galaxy.
Our calculations highlight the importance of cosmic rays in these measures,
making them an essential ingredient of realistic models
of the interstellar medium.
\end{abstract}

\title{The Parker instability in disk galaxies}
\author{L.~F.~S.~Rodrigues,\altaffilmark{1}
  G.~R.~Sarson,\altaffilmark{1} A.~Shukurov,\altaffilmark{1} P.~J.~Bushby,\altaffilmark{1}
  and A.~Fletcher\altaffilmark{1}
}
\altaffiltext{1}{\href{mailto:luiz.rodrigues@newcastle.ac.uk}{
                      luiz.rodrigues@newcastle.ac.uk (LFSR)},\hfill\\
                 \href{mailto:graeme.sarson@newcastle.ac.uk}{
                       graeme.sarson@newcastle.ac.uk (GRS)},\hfill\\
                 \href{mailto:anvar.shukurov@newcastle.ac.uk}{
                       anvar.shukurov@newcastle.ac.uk (AS)},\hfill\\
                 \href{mailto:paul.bushby@newcastle.ac.uk}{
                       paul.bushby@newcastle.ac.uk (PJB)},\hfill\\
                 \href{mailto:andrew.fletcher@newcastle.ac.uk}{
                       andrew.fletcher@newcastle.ac.uk (AF)}.\hfill\\
                }
\affil{School of Mathematics and Statistics, Newcastle University,
Newcastle upon Tyne, NE1 7RU, UK}


\section{Introduction}

The magnetic buoyancy (or magnetic Rayleigh--Taylor) instability 
enhanced by cosmic rays in disk galaxies is known as
the Parker instability \citep{Parker1966,Parker1967,Parker1969}. Since cosmic rays
are practically weightless but exert pressure comparable to that of thermal gas,
turbulence and magnetic fields, their effect on the instability is significant.
The growth rate of the most unstable mode is of the order of the Alfv\'en crossing time,
$\tau\A= h/c\A\simeq10^7\yr$, \revision{based on the disk scale height} $h=100\pc$, \revision{and assuming an Alfv\'en speed of} $c\A=10\kms$; the corresponding wavelength is of order $1\kpc$ \citep{P79}.
The nature of the instability is quite generic and it is therefore expected 
to occur in \revision{all} magnetized astrophysical disks, 
including accretion disks and spiral galaxies.
The saturation mechanism of the instability and the resulting state of the
system are not well understood. 
As discussed by \citet{FT94,FT95}, differential rotation weakens the instability. 
Gas viscosity, random magnetic fields and finite cosmic ray diffusion 
\revision{also} suppress it \citep{KP83},
as they interfere with the sliding of the gas to the
bottom of magnetic loops as well as the streaming of cosmic rays to the top of these loops.

There have been several studies of the Parker instability using linear perturbation
theory.
\citet{Giz1993} extended the initial works by Parker by employing a more realistic
vertical gravitational acceleration, $g_z(z)\propto\tanh{z}$,  
which is continuous across the Galactic midplane, 
rather than the step-function $|g_z(z)|=\mathrm{constant}$ 
that was previously used. 
This allowed them to consider the vertical symmetry of the 
instability and identify a family of unstable modes that cross the midplane \revision{(in which case the vertical velocity, $U_z$, and the vertical magnetic field, $B_z$, are not identically zero at $z=0$). Modes of this type, which are often referred to as `antisymmetric', were also found by \citet{Horiuchi1988}, although they used a different gravitational acceleration, $g(z)\propto z/(r_0^2+z^2)^{3/2}$.} 
\revision{(The issue of symmetry about the midplane 
is discussed in detail in section~\ref{sec:depz}.)}
\citet{Kim1998} showed that the \revision{use of a} more realistic gravity \revision{profile} also 
results in a faster growth rate and shorter horizontal wavelength for the instability.  
\citet{Ryu2003} included the effects of finite cosmic ray diffusion;
the previous works had taken the diffusivity to be infinite 
along the magnetic field and zero perpendicular to it. 
Their linear analysis found that the finite, but large, 
parallel diffusivity of cosmic rays in the interstellar medium (ISM) 
is well approximated by infinite-diffusivity models, 
and that the perpendicular diffusivity has no significant effect.  
\citet{Kuwabara:2006} studied Parker-Jeans instabilities in the ISM using a model which 
does not have an external gravitational field but does include self-gravity. They confirmed 
that higher cosmic ray diffusivity results in faster growth rates,
and that modes with a wavevector  
parallel to the magnetic field direction are least stable.

There have also been many numerical studies of the Parker instability. 
To date most numerical models have been ideal:
the terms involving the gas viscosity in the Navier-Stokes equation, 
and the magnetic diffusivity in the magnetic induction equation, 
are set to zero\revision{;
notable exceptions are \citet{Hanasz2002} and
\citet{Tanuma2003}, which did include finite resistivity. Numerical}
models have generally assumed that the vertical gravity is uniform 
and discontinuous at the midplane and/or do not include cosmic rays. 
\citet{Matsumoto1988} and \citet{Basu1997} used 2D simulations to investigate the
nonlinear evolution of the instability,
and showed that the midplane crossing modes dominate. 
\citet{Kim:1998b} found that filamentary or sheetlike structures formed when the system 
approaches a new equilibrium after $10^{9}\yr$ in 3D models.
Disk rotation was included in
the 3D simulations of \revision{\citet{Chou1997} and} \citet{Kim:2001},
who showed that the Coriolis force results in a twisting
of the magnetic loops and may also help to randomize the initially uniform magnetic field. 3D 
models including self-gravity and differential rotation, allowing for the development of Jeans
and swing instabilities as well as the Parker instability, were used by \citet{Kim2002} to show that a mixture
of midplane symmetric and midplane crossing modes can develop in the ISM. The above simulations
all excluded a cosmic ray component, which was added by \citet{Hanasz:2000, HanaszLesch2003} in
3D, to investigate the triggering of the Parker instability by a supernova remnant, and by \citet{Kuwabara:2004},
who used 2D MHD simulations to confirm the connection between cosmic ray diffusivity and the Parker instability
growth rate found in their earlier linear analysis. \citet{Mouschovias2009} allowed for phase
transitions in a 2D MHD model (without cosmic rays), found that the midplane crossing mode was
favoured, and argued that a non-linear triggering of the Parker instability by a spiral density shock-front could
allow the instability to form gas clouds on a timescale compatible with observations.

The instability produces buoyant loops of a large-scale magnetic field 
at a kiloparsec scale.
These `Parker loops' are expected to lie largely in the azimuthal direction
(the direction of the large-scale field).
This corresponds to the `undular' modes 
(with wavevector parallel to magnetic field $\bm{B}$)
which are expected to dominate over the `interchange' modes 
(with wavevector perpendicular to $\bm{B}$) in most linear analyses 
\citep[see, e.g.][for a discussion of these modes]{Matsumoto1993}.
(Indeed, some authors identify the Parker instability 
specifically with these undular modes,
rather than simply in terms of the fundamental buoyancy mechanism.)
There have been numerous attempts to detect such Parker loops
in observations, but they have met with little success\revision{\footnote{
\revision{
Recently, magnetic loops and filaments have been
detected in WMAP and Planck data \citep{Vidal2015,Planck2015}. However, these are
likely a consequence of interstellar turbulence driven by supernovae remnants
\citep{Planck2015,PlanckExplanation,Mertsch2013} and not of Parker instability.
}}}.
This may be surprising, since the instability is strong and generic. 
Therefore, we consider here the detailed development and possible 
observational consequences of the instability in galactic disks.

We simulate the evolution of a simple model of a section of a galactic disk, where
gas, magnetic fields and cosmic rays evolve in the gravitational field of stars and
dark matter halo. Cosmic rays are described in the advection-diffusion approximation
with a diffusion tensor aligned with the local magnetic field.
The simulation setup is aimed at facilitating comparisons 
with previous analytical works, 
as well as capturing clear signatures of the Parker instability 
which can later be compared to more advanced simulations and observations. 
For these reasons, we use the isothermal approximation
(where there is no possibility of confusion 
with thermally-driven instabilities),
and omit rotation and rotational shear
(to avoid any confusion with other instabilities,
such as the magneto-rotational instability).
We also refrain from using any source terms for the magnetic field 
and the cosmic ray energy density,
since this could lead to artificial spatial biases.

This paper is organized as follows. In Section \ref{sec:model} the model is
described in detail. In Section \ref{sec:results} the results are presented:
in \S\ref{sec:alphabeta} we find the growth rates and typical wavenumbers
associated with the instability, 
as well as the dependence on the density of cosmic rays 
and the strength of the magnetic field; 
in \S\ref{sec:depz} we examine the symmetries of the instability,
and the variation of the growth rates with height;
in \S\ref{sec:Pm} we address the dependence on the choice of diffusivities;
and in \S\ref{sec:obs} we study possible observable signatures: namely the
prospects of observing Parker loops from synchrotron emission of edge-on
galaxies, \revision{or from} the Faraday rotation measure signal from face-on galaxies. 
Finally, in Section \ref{sec:conc} we state our conclusions.

\section{Model description}
\label{sec:model}

All the numerical calculations in this work were performed using the open
source, high-order finite difference \pencil code\footnote{Documentation and 
download instructions can be found at \url{http://pencil-code.nordita.org/} .}, 
designed for fully non-linear,
compressible magnetohydrodynamic simulations. The cosmic ray calculations employed
the \textsc{cosmicray} and \textsc{cosmicrayflux} modules \citep{Snodin2006}.

\subsection{Basic equations and cosmic ray modeling}
\label{sec:equations}
We solve equations for mass conservation, momentum, and magnetic induction,
for an isothermal gas assuming constant kinematic viscosity $\nu$
and magnetic diffusivity $\eta$:
\begin{align}
 \frac{\Dv\ln\rho}{\Dv t}& = -\nabla\cdot\bm{U}\,,\\
 \frac{\Dv \bm{U}}{\Dv t} &= \bm{g} -\frac{\nabla(p_{\rm g}+p\CR)}{\rho} +
\frac{\bm{j} \times \bm{B}}{\rho} \nonumber \\ 
  &+ \nu\left[\nabla^2\bm{U}
		+\tfrac{1}{3}\nabla\left(\nabla\cdot\bm{U}\right)
			+2\mathbb{W}\cdot{\nabla}\ln\rho \right]\,, \label{eq:mom} \\
\frac{\partial\bm{A}}{\partial t} &= \bm{U}\times\bm{B} +\eta\nabla^2\bm{A} \,,
\end{align}
where $\rho$ is the gas density,
$\bm{U}$ is the gas velocity,
$\bm{A}$ is the magnetic vector potential,
$\bm{B}=\nabla\times\bm{A}$ is the magnetic field
(with $\nabla \cdot \bm{B}=0$ by construction),
$\bm{j} = \nabla\times\bm{B}\,(c/4\pi)$ is the electric current density, 
\revision{with $c$ the speed of light in vacuum,}
$\mathbb{W}$ is the rate of strain tensor,
\[
    W_{ij}=\frac{\partial U_i}{\partial x_j}+\frac{\partial U_j}{\partial x_s}
    - \frac{2}{3}\delta_{ij}\nabla_k U_k \,,
\]
$p_{\rm g}=\rho c_{\rm s}^2$ is the gas pressure,
$c_{\rm s}$ is the adiabatic speed of sound,
$p\CR$ is the cosmic ray pressure and $\bm{g}$ is the gravitational acceleration.
We neglect the effects of rotation
and rotational shear.

Cosmic rays are modeled using a fluid approximation
\citep[e.g.][]{Parker1969,Schlickeiser1985}
where the cosmic ray energy density $\ecr$ is governed by
\begin{equation}
\label{eq:ecr}
 \frac{\partial\ecr}{\partial t}+\nabla\cdot(\ecr\bm{U})+p\CR\nabla\cdot\bm{U}=
 -\nabla\cdot\vect{\Fcrv}\, ,
\end{equation}
with $\vect{\Fcrv}$ the cosmic ray flux defined below.
The cosmic ray pressure that appears in equation~\eqref{eq:mom} is obtained from the equation of state
\[
p\CR=\epsilon\CR(\gamma\CR-1)\,,
\]
with $\gamma\CR$ the cosmic ray adiabatic index. We adopt $\gamma\CR=4/3$,
as appropriate for the ultrarelativistic gas;
other values in the range $4/3 \le \gamma\CR \le 5/3$ are possible,
with the latter value being relevant for non-relativistic particles
\citep{Schlickeiser1985}.

The cosmic ray flux $\vect{\Fcrv}$ is introduced in a non-Fickian
form justified and discussed by \citet{Snodin2006},
\begin{equation}
\label{eq:fcr}
 \tau \frac{\partial \Fcrv_i}{\partial t}=
-\kappa_{ij}\nabla_j\ecr
 - \Fcrv_i\, ,
\end{equation}
where $\tau$
can be identified with the decorrelation time of the cosmic ray pitch angles,
and $\kappa$ is the diffusion tensor,
\begin{equation}
 \kappa_{ij}=\kappa_\perp\delta_{ij}+(\kappa_\parallel-\kappa_\perp)\hat{B}_i
\hat{B}_j\,,
\end{equation}
where a circumflex denotes a unit vector.
When $\tau$ is negligible, equation~\eqref{eq:fcr} reduces to the Fickian diffusion flux,
but the large diffusivity of cosmic rays makes the non-Fickian effects both physically
significant and numerically useful. When $\tau$ is finite,
the propagation speed of cosmic rays remains finite \citep[e.g.,][]{Bakunin}.
This prescription also avoids the possibility of singularities
at X-type magnetic null points
where $\kappa_{ij}$ is undefined \citep{Snodin2006}.

\revision{For numerical solution, 
the maximum time step $\delta t$ for the cosmic ray physics
might be constrained either by an advective-style Courant condition 
based on the cosmic ray flux equation (\ref{eq:fcr}),
or by a diffusive-style Courant condition based on the cosmic ray energy
density equation (\ref{eq:ecr}). 
(The latter would be appropriate for the Fickian case $\tau=0$.)
I.e.\ we might constrain $\delta t \le C_{\rm adv} \delta z /c\CR$, 
or $\delta t \le C_{\rm diff} \delta z^{2} /\kappa_\parallel$,
with $\delta z$ the smallest numerical grid spacing,
and $c\CR=(\kappa_{\parallel}/\tau)^{1/2}$
the characteristic speed for the cosmic ray diffusion 
along the local magnetic field;
here $C_{\rm adv}$ and $C_{\rm diff}$ are empirical constants.
For similar advective and diffusive processes in the other equations
(based on speeds $c_{\rm s}$, $c_{\rm A}$, $U$,
and diffusivities $\nu$, $\eta$, respectively),
we adopt the constants $C_{\rm adv}=0.4$, $C_{\rm diff}=0.06$.
For the relatively large value of $\kappa_\parallel$ used in the following,
this value of $C_{\rm diff}$ results in a significant reduction of $\delta t$,
to a value which numerical experiments show to be unnecessarily small.
I.e.\ an appropriate empirical value of $C_{\rm diff}$ for the cosmic ray 
diffusion would be significantly greater than our standard value of 0.06.
(This is perhaps unsurprising, given the anisotropic nature 
of our cosmic ray diffusion;  
$\kappa_\parallel$ only applies along magnetic field lines,
where cosmic ray energy density gradients tend to be small.) 
On the other hand, our standard value of $C_{\rm adv}=0.4$ 
results in an appropriate constraint for $\delta t$
(allowing the cosmic ray physics to control the timestep when necessary,
but avoiding unnecessarily small timesteps),
and achieves this without the introduction of an additional empirical constant.
In the calculations reported here, we use this advective-style constraint.}

\subsection{Numerical domain and initial conditions}
\label{sec:initial}

We use a local Cartesian box to represent a portion of the galactic disk,
with $x$, $y$ and $z$ representing
the radial, azimuthal and vertical directions, respectively.
The domain is of size $6\kpc\times12\kpc\times3.5\kpc$
in the $x$, $y$ and $z$ directions,
with the galactic midplane at the center of the vertical range
$-1.75\kpc \le z \le 1.75\kpc$.
We use a grid of $256\times512\times315$ mesh points,
so have the numerical grid spacings
$(\delta_x,\delta_y,\delta_z)\approx(23.4\,\pc,23.4\,\pc,
11.1\,\pc)$.
The height of the domain has been chosen to be much larger
than the characteristic vertical scale of the instability (see below).

The initial state is an isothermal disk under \revision{magnetohydrostatic}
equilibrium in an external gravity field with the gravitational acceleration
given by
\begin{equation}
  g_z(z) = -2\pi G \Sigma \tanh\left(\frac{z}{h}\right)\,,
\end{equation}
where $h=500\pc$ is the the scale height of warm gas,
$\Sigma=100 \msun \pc^{-2}$ is the total surface mass density of the disk,
and $G$ is Newton's gravitational constant.

The relative contributions of the gas, magnetic and cosmic ray to the total
pressure in the initial state are parameterized using
\begin{equation}
\label{eq:alphabeta}
  \alpha = \frac{p_B}{p_\text{g}} \quad \text{and}\quad
  \beta = \frac{p\CR}{p_\text{g}} \text{ ,}
\end{equation}
where $p_B=B^2/(8\pi)$ is the magnetic pressure.
The initial state is unstable for any values $\alpha+\beta>\gamma_{\rm g}-1=0$ 
(where $\gamma_{\rm g}=1$ is the adiabatic index of the gas).

The
temperature of the gas follows as
\begin{equation}
T = \frac{\pi G \,\Sigma m_\mathrm{p} \mu h}{k_\text{B} (1+\alpha+\beta)}\,,
\end{equation}
where $m_\mathrm{p}$ is the proton mass,
$\mu$ is the mean molecular mass (here $\mu=1$),
and $k_\text{B}$
is Boltzmann's constant.
Under \revision{magnetohydrostatic} equilibrium with the gravity field adopted, the initial gas
density is given by
\begin{equation}
  \rho(z) =  \frac{\Sigma\gas}{2\,h}\text{sech}^2\left(\frac{z}{h}\right)\,,
\end{equation}
where $\Sigma\gas$ is the gas surface density.

When the pressure scale heights of magnetic field and cosmic rays are
identical, as imposed by equation~\eqref{eq:alphabeta},
the initial energy density distribution of cosmic rays is
\begin{equation}
 \epsilon\CR(z) = \frac{\pi G\Sigma \Sigma_{\rm g}}{2 (\gamma\CR-1)}
\text{sech}^2\left(\frac{z}{h}\right)
               \frac{\beta}{1+\alpha+\beta}\,.
\end{equation}
The initial magnetic field, assumed to be azimuthal, is obtained from
\begin{equation}
\label{eq:initA}
\begin{split}
        A_x &= 4 \pi h \left(
              \frac{ G \Sigma\Sigma\gas\alpha}{1+\alpha+\beta}\right)^{1/2}
               \arctan\left[\tanh\left(\frac{z}{2h}\right)\right]
               \,,\\
        A_y &= A_z = 0 \,,
\end{split}
\end{equation}
\revision{which leads to}
\begin{align}
	B_y^2 &= 4\pi^2 G\Sigma\Sigma_{\rm g} \text{sech}^2\left(\frac{z}{h}\right)
               \frac{\alpha}{1+\alpha+\beta}
               \,, \nonumber\\
        B_x^2 &= B_z^2 = 0 \,.
\end{align}

This initial state is not in exact \revision{magnetohydrostatic} equilibrium, except for 
$\alpha=\beta=0$,
due to the magnetic diffusion and perpendicular cosmic ray diffusion.
For the parameters used, the diffusive time scales are much longer
than the timescale of the Parker instability, so any \revision{diffusive} evolution of the initial state
is inconsequential (see Section~\ref{sec:timescales}).

\begin{deluxetable*}{lcc}
\tablecaption{Various timescales of the problem, with respect to 
              vertical disk crossing times.\label{tab:timescales}}
\tablecolumns{3}
\tablehead{\colhead{Time scale} & \colhead{Definition}  & \colhead{Magnitude}}
\startdata
Sound speed   &
$\displaystyle \tau_s =
\frac{h}{c_s}=\left(\frac{h}{\gamma_{\rm
g}\pi G\Sigma}\right)^{1/2} (1+\alpha+\beta)^{1/2}$
& $\displaystyle 26.6\Myr\left(\frac{1+\alpha+\beta}{2}\right)^{1/2}$\\[1em]
Alfv\'{e}n speed \footnote{Initial value.}
& $\displaystyle \tau_A = \frac{h}{c_A} = \left(\frac{h}{2\pi
G\Sigma}\right)^{1/2}\left(
\frac{1+\alpha+\beta}{\alpha}\right)^{1/2}$  & $\displaystyle 18.8\Myr\left(
\frac{1+\alpha+\beta}{2\alpha}\right)^{1/2}$\\[1em]
Viscous  & $\tau_\nu = \displaystyle\frac{h^2}{\nu}$  & $2.39\Gyr$\\[1em]
Magnetic diffusion & $\tau_\eta = \displaystyle\frac{h^2}{\eta}$  & $ 4.77\Gyr$\\[1em]
Cosmic ray parallel diffusion & $\tau_{\text{cr},\parallel} =
\displaystyle\frac{h^2}{\kappa_\parallel}$  & $ 4.77\Myr$\\[1em]
Cosmic ray perpendicular diffusion & $\tau_{\text{cr},\perp} = \displaystyle\frac{h^2}{\kappa_\perp}$
& $2.39\Gyr$\\[1em]
Cosmic ray flux propagation speed\footnote{Based on parallel diffusion.}
& $\tau_{\text{cr},\tau} = \displaystyle\frac{h}{c\CR} = \displaystyle h\,/\sqrt{ {\kappa_\parallel}/{\tau} }$  & $6.91\Myr$
\enddata
\end{deluxetable*}

\begin{deluxetable}{lcc}
\tablecaption{Values of parameters adopted\label{tab:diffusivities}}
\tablecolumns{3}
\tablehead{\colhead{Quantity} & \colhead{Symbol}  & \colhead{Value}}
\startdata
Disk gas surface density         & $\Sigma_{\rm g}$       &
$10\msun\pc^{-2}$\\[0.85ex]
Disk total surface density      & $\Sigma$    & $100\msun\pc^{-2}$\\[0.85ex]
Disk scale-height         & $h$         & $0.5\kpc$\\[0.85ex]
CR flux correlation time  & $\tau$             &  $10\Myr$\\[0.85ex]
Kinematic viscosity          & $\nu$ & $3.16\times 10^{25}\cm^2\s^{-1}$ \\[0.85ex]
Magnetic diffusivity         & $\eta$             & $1.58\times 10^{25}\cm^2\s^{-1}$ \\[0.85ex]
CR parallel diffusivity      & $\kappa_\parallel$ & $1.58\times 10^{28}\cm^2\s^{-1}$\\[0.85ex]
CR perpendicular diffusivity & $k_\perp$ & $3.16\times 10^{25}\cm^2\s^{-1}$
\enddata
\end{deluxetable}

The initial state is perturbed by noise in the
velocity field: at each mesh point a random velocity was specified
with a magnitude uniformly distributed in the range
$0\leq U\leq 14\, \text{sech}^2(z/h)\kms$.
The velocity perturbation is subsonic, with the rms value of order
$1\kms$ throughout the domain. 
This value was chosen to avoid unnecessarily long transients in the onset
of the instability
(during which time the initial state might evolve significantly),
while avoiding directly driving the instability.

\subsection{Boundary conditions}

We assume periodicity at the horizontal ($x$ and $y$) boundaries.
The height of the domain, $Z_0=1.75\kpc$, is much larger than
the characteristic vertical scale of the instability,
so that the precise boundary conditions at the top and bottom of the domain
do not significantly affect the results; these are adopted as follows.
Symmetry about the top and bottom boundaries is imposed for $U_x$ and $U_y$,
and antisymmetry for $U_z$:
\begin{equation}
\frac{\partial U_x}{\partial z} =\frac{\partial U_y}{\partial z}= U_z = 0
\quad\text{at }
|z|=Z_0\,,
\end{equation}
making these boundaries impenetrable and stress free.
The magnetic vector potential is symmetric for $A_z$,
antisymmetric for $A_y$ and generalized antisymmetric for $A_x$:
\begin{equation}
\frac{\partial^2 A_x}{\partial z^2}
=A_y
=\frac{\partial A_z}{\partial z}= 0
\quad\text{at }
|z|=Z_0\,.
\end{equation}
These conditions are similar to those for a perfectly conducting boundary
(which would require $A_x=0$),
but the condition of antisymmetry in $A_x$ has been relaxed
to avoid overly distorting the initial background profile $A_x(z)$.
Similarly, to minimize distortions to the initial background density profile,
generalized antisymmetry is imposed on $\ln\rho$ at the top and bottom of the domain:
\begin{equation}
\frac{\partial^2 \ln\rho}{\partial z^2}=0
\quad\text{at }
|z|=Z_0\, .
\end{equation}
This condition has no obvious physical interpretation but
it places no strong constraints on the density distribution.
The cosmic ray energy density and flux at the
top and bottom boundaries satisfy
\begin{equation}
 \frac{\partial \epsilon\CR}{\partial z}=0 \,, \qquad
 \frac{\partial \Fcrv_x}{\partial z}=
 \frac{\partial \Fcrv_y}{\partial z}=
 \Fcrv_z = 0\,,
\end{equation}
which corresponds to vanishing vertical flux in the Fickian limit.

The boundary conditions have been chosen largely for simplicity.
Because of the impenetrability of the surface of the domain, 
there could in principle be reflection of waves 
and/or the artificial accumulation of buoyant material,
but no such problems emerged in the results reported below (which focus primarily
upon the early phases of evolution of the Parker instability).
\revision{We have also rerun one of our simulations in a domain with larger vertical
size to confirm that the domain is large enough to prevent spurious boundary
effects over the timescales explored (see appendix \ref{ap:tall_box})}.

\subsection{Parameters and timescales}
\label{sec:timescales}

The parameters used,
including our fiducial choices of the various diffusivities involved,
are shown in table~\ref{tab:diffusivities};
the resulting timescales of interest for our disk
are shown in table~\ref{tab:timescales}.
These values of $\Sigma$ and $\Sigma_{\rm g}$ correspond to an initial midplane
gas density of $\rho(0)=6.8\times 10^{-25} \; {\rm g\,cm}^{-3}$;
for $\alpha=1$, $\beta=1$,
the corresponding initial azimuthal field strength
is
$B_{y}(0)\approx 6\; {\rm \mu G}$
and the
cosmic ray energy density,
$\epsilon\CR(0)~\!\approx~\!4.6\times 10^{-13} \; {\rm erg\,cm}^{-3}$.

Our choice of $\tau=10\Myr$ for the cosmic ray flux correlation timescale
ensures that it is smaller than the characteristic timescale
of the Parker instability (see Section~\ref{sec:results}),
as well as the sound and Alfv\'{e}n speed timescales
(see table~\ref{tab:timescales}).
The impact of choosing smaller values for $\tau$ was examined,
and confirmed to have negligible impact on the results.
This value of $\tau$, together with our other choices of parameters,
results in the Strouhal number associated with the cosmic ray flux,
$\text{St} = c\CR \tau/h= \sqrt{\kappa_\parallel \tau}\,/h \approx 1.45$,
being of order 1.
(And note that this estimate, involving parallel diffusivity with the
vertical scale height, is arguably an overestimate.
Typical horizontal lengthscales are longer than $h$;
see Section~\ref{sec:alphabeta}, below.)
Thus we do not expect wavelike behaviour in these simulations;
the diffusion of cosmic rays does not differ significantly
from Fickian diffusion.

Our fiducial choices of kinematic viscosity $\nu$
and magnetic diffusivity $\eta$
correspond to a magnetic Prandtl number $\text{Pm}=\nu/\eta=2$;
the effects of varying magnetic Prandtl number are investigated in
Section~\ref{sec:Pm}.
The values adopted for $\nu$ and $\eta$ are very significantly larger
than expected molecular values
(and the resulting $\text{Pm} \sim {\cal O}(1)$ is very much smaller
than the galactic regime associated with molecular values, $\text{Pm}\gg1$);
but if considered as effective turbulent diffusivities,
the values are reasonable.

The parallel cosmic ray diffusivity used is close to the best estimates
for this parameter \citep{Ryu2003,StrongMoskalenko1998}.
The timescales associated with cosmic ray diffusion along the fieldlines
($\tau_{{\rm cr},\parallel}$ and $\tau_{{\rm cr},\tau}$
in table~\ref{tab:timescales})
are significantly shorter than the timescales of the Parker instability,
and the sound and Alfv\'{e}n speed timescales;
we are therefore in a reasonable regime for modelling galactic disks,
yet may expect some deviations from those analytic results
obtained on the basis of infinitely fast propagation of cosmic rays
along fieldlines.

The perpendicular diffusivity of cosmic rays, on the other hand,
was somewhat reduced from expected values to avoid the escape of
the cosmic ray energy density before the development of the instability.
As noted in Section~\ref{sec:initial}, the initial state will evolve under
the diffusion of magnetic field and the perpendicular diffusion of cosmic rays;
but the timescales for these processes, 4.77 Gyr and 2.39 Gyr,
are very long compared to the other timescales of interest,
making the slow diffusive evolution of the background state insignificant.

\begin{figure*}
\centering
 \includegraphics[width=0.9\textwidth]{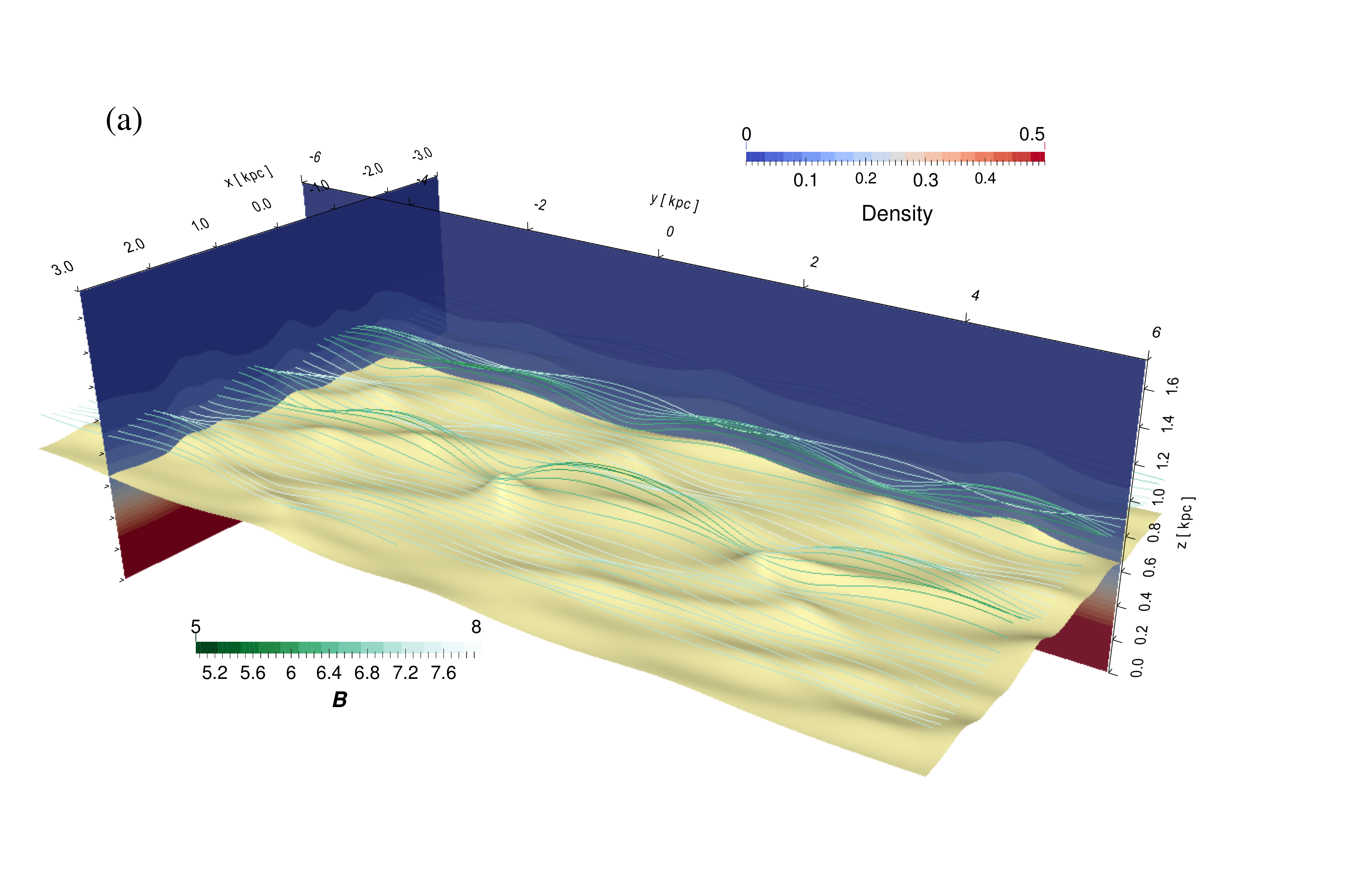}
 \includegraphics[width=0.9\textwidth]{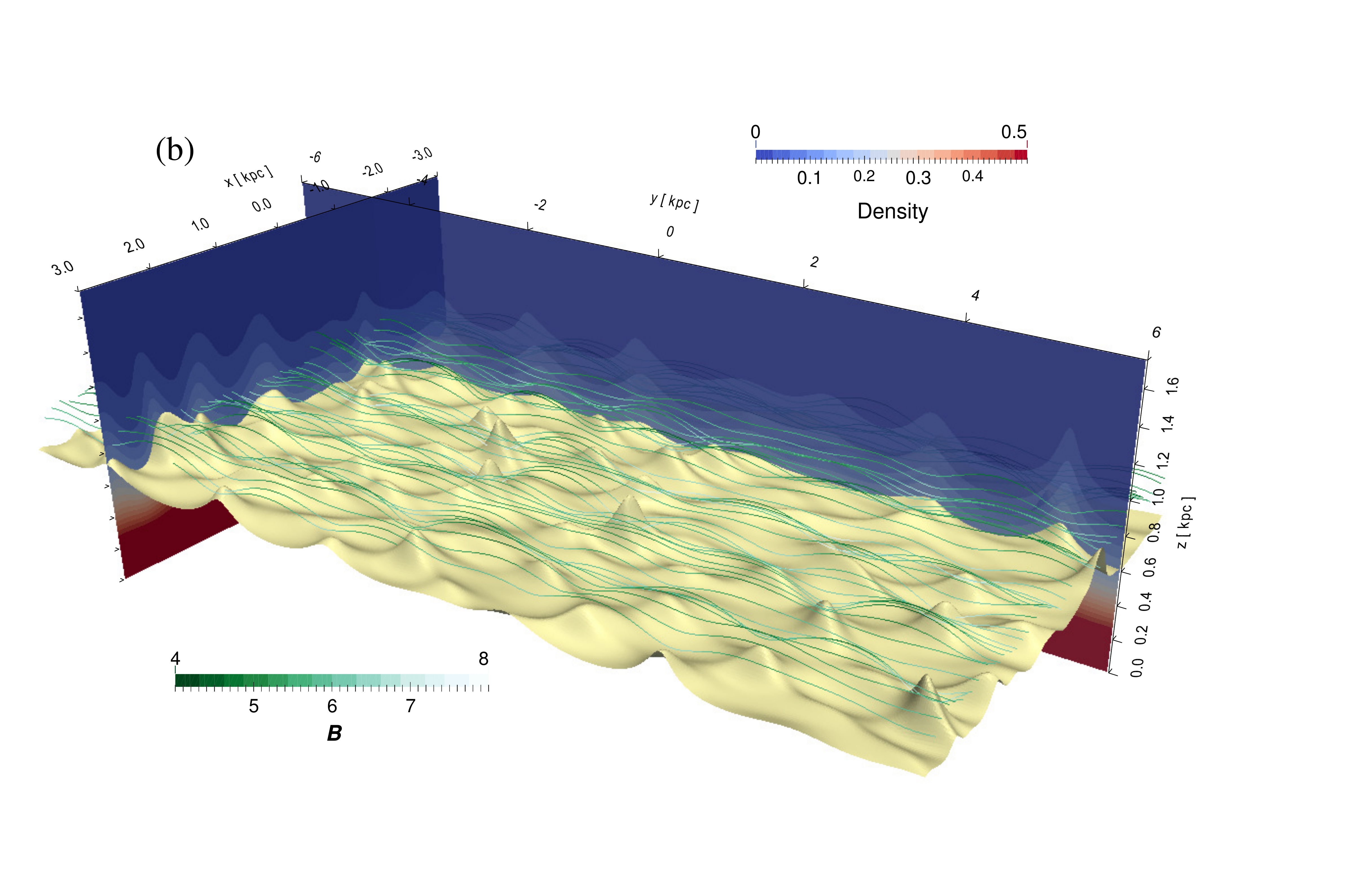}
 \caption{
          Renderings of the simulations at the time where the kinetic energy
          is close to the thermal energy.
          Panel (a) shows the outcome of the choice
          $\alpha=1.0$ and $\beta=0.0$ at $t=330\Myr$.
          Panel (b) shows the outcome of the choice
          $\alpha=1.0$ and $\beta=1.0$ at $t=200\Myr$.
          The yellow surface corresponds to a density of
          $1.5\times10^{-25} \g \cm^{-3}$.
          The green lines represent magnetic field lines, color-coded by field
          magnitude following the color-bar in the plots,
          which is units of $0.35\muG$.
          The slices show the density in units of $10^{-24}\g \cm^{-3}$.
          Only the top half (i.e. the volume between the midplane and the top
          boundary) of the simulation domain is shown.
          }
 \label{fig:pretty}
\end{figure*}

\section{Results and discussion}
\label{sec:results}

Three-dimensional renderings of the runs of the simulation with
$(\alpha,\beta)=(1,0)$ and $(\alpha,\beta)=(1,1)$ are shown in
Figure~\ref{fig:pretty}, in panels (a) and (b) respectively.
The images show the snapshots when the kinetic energy in each simulation
matched the thermal energy (which is fixed due to the isothermal condition);
this corresponds to simulation time $330\Myr$ for the case without cosmic rays
(panel a)
and $200\Myr$ for the case of equal pressure contribution of cosmic rays,
magnetic fields and thermal pressure (panel b).

As expected (and as quantified further below),
one can see that the presence of cosmic rays leads to an
increased number of substructures in the density field.
The magnetic field lines, in green, display the characteristic `Parker loops',
which also clearly span smaller length scales in the case with cosmic rays.

\subsection{Dependence on magnetic and cosmic ray pressures ($\alpha$ and $\beta$)}
\label{sec:alphabeta}

\begin{figure}
 \includegraphics[width=\columnwidth]
 {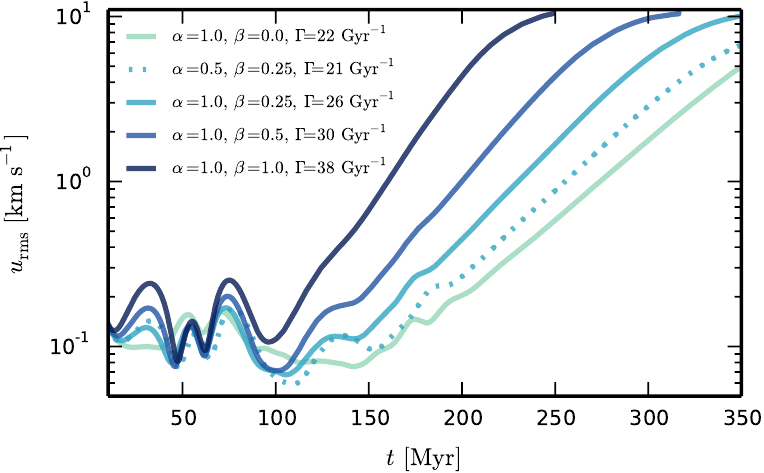}
 \includegraphics[width=\columnwidth]
 {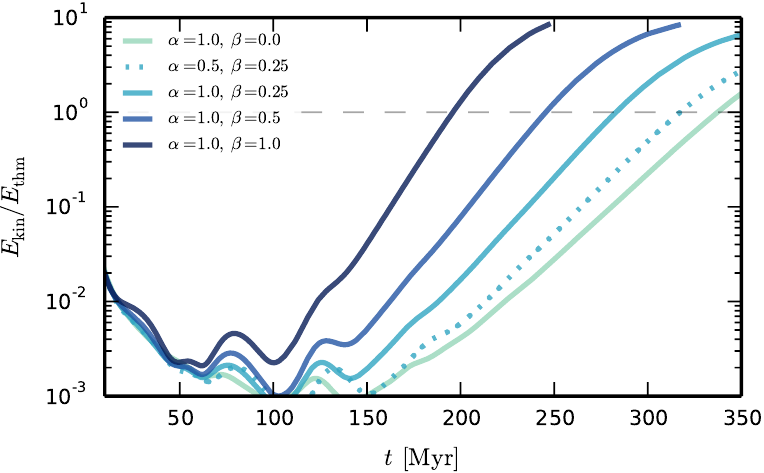}
 \caption{
          The top panel shows the evolution of the the root mean square of
          the velocity, $U_{\rm rms}$, 
          for selected choices of $\alpha$ and $\beta$ as indicated.
          The bottom panel shows the evolution of the kinetic energy in the
          whole simulation box (normalised by the fixed thermal energy). The horizontal dashed line corresponds to
          equipartition of thermal and kinetic energy;  the analysis in the
          text is for states before this point,
          when $E_{\rm kin}=0.7 E_{\rm therm}$.
         }
 \label{fig:growth_rate}
\end{figure}

In Figure \ref{fig:growth_rate},
the time evolution of the root-mean-square velocity $U_{\rm rms}$,
and of the kinetic energy $E_{\rm kin}$,
are shown for different choices of $\alpha$ and $\beta$.
A period of exponential growth, associated with the development of the
instability, can be clearly identified.
Increases in the pressure ratio of magnetic fields,
$\alpha$, and of cosmic rays, $\beta$, enhance the instability,
leading to larger growth rates.

\begin{figure}\centering
 \includegraphics{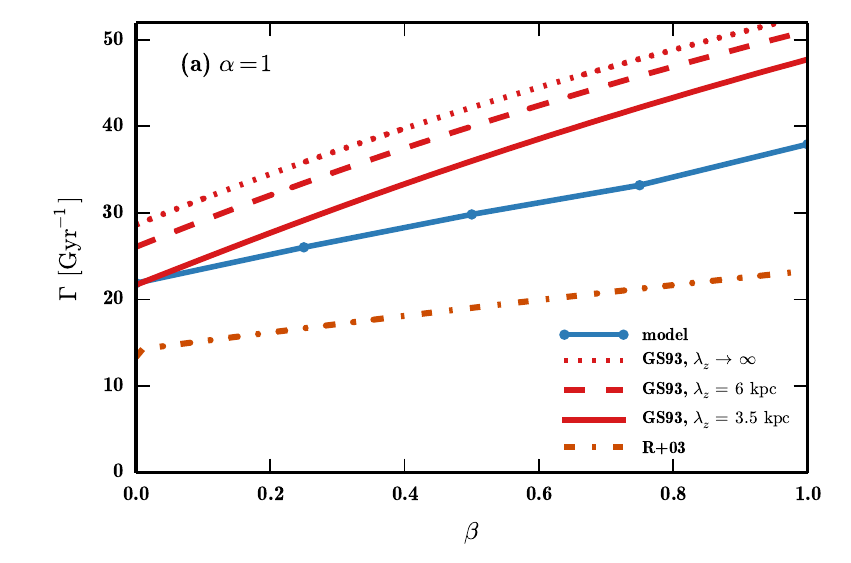}
 \includegraphics{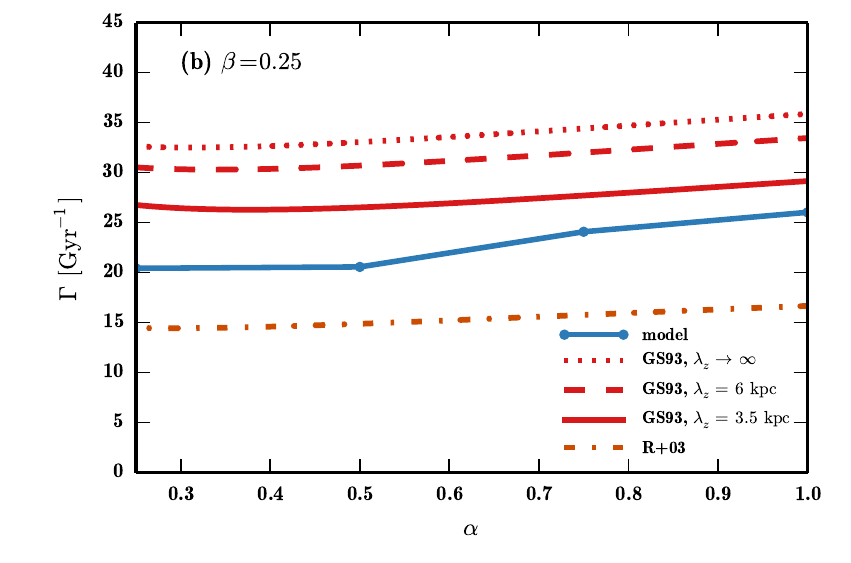}
 \caption{
             Dependence of the growth rate of the velocity
             field with pressure ratio.
             Panel (a): varying cosmic ray pressure ratio, $\beta$
             (with fixed $\alpha=1$), is shown by solid blue curve.
             Panel (b): magnetic field pressure ratio, $\alpha$ (with
             fixed $\beta=0.25$), is shown by solid blue curve.
             In both panels, the fastest growing modes \revision{
             computed using the equations derived by \citet[][GS93]{Giz1993} 
             correspond to red curves. Different choices for the vertical 
             wavelengths, $\lambda_z$, are shown with different linestyles,
             as indicated.
             The fastest growing modes predicted by \citet[][R+03]{Ryu2003} are
             shown by the bottom dash-dotted curve.}
         }
 \label{fig:growth}
\end{figure}

The dependence of the growth rate of $U_z$, $\Gamma$, on $\alpha$ and $\beta$
is examined in Figure \ref{fig:growth}, where it is compared with the analytical
results of \citet[][GS93]{Giz1993} and \citet[][R+03]{Ryu2003}.
Under the assumption of equipartition of initial magnetic pressure ($\alpha=1$),
the growth rate increases almost linearly with cosmic ray pressure contribution,
from $20.4\Gyr^{-1}$, at $\beta=0$, to $26.0\Gyr^{-1}$, at $\beta=1$. \revision{
To compare our findings to previous analytical results, one has to take into account
the effect of a finite vertical domain with impenetrable boundaries, which limits 
the maximum vertical wavelength of the perturbation, $\lambda_z$. 
In order to illustrate this,  we show three sets of analytic growth rate curves for different choices of 
$\lambda_z$. 
For arbitrarily large vertical wavelengths GS93 predict 
fastest growing modes with growth rates
$28.6\Gyr^{-1}$ and $52.8\Gyr^{-1}$ at $\beta=0,1$, respectively, which differ
significantly from our results. 
If, instead, the maximum attainable vertical wavelength in our domain is 
considered, $\lambda_z=3.5\kpc$, the growth rates are closer to what is obtained
in the simulations. For the $\beta=0$ case (i.e. the run without 
cosmic rays), there is good quantitative agreement between the model and GS93. However, the 
slope, $\dv \Gamma /\dv \beta$, found from the simulation results is clearly 
smaller. 
It should be noted that GS93
assumes infinite diffusion of cosmic rays parallel to the field and 
no perpendicular diffusion,
and both of these factors would lead us to expect smaller growth rates
\citep{Ryu2003}.
}

The analytical results of \citet{Ryu2003},
which take into account finite cosmic ray diffusivity,
give a better match for the slope \revision{$\dv \Gamma /\dv \beta$}, 
but have smaller growth rates
because of the assumption of a constant gravitational field;
the importance of the gravity profile in this respect is clear from
\citet{Kim1998}, who considered uniform, linear and `realistic' fields
(the latter being similar to our own).

In the lower panel of Figure~\ref{fig:growth},
the weaker dependence of the growth rate on $\alpha$ is shown,
for a fixed $\beta=0.25$.
This relatively weak dependence is expected for realistic gravity fields
\citep{Kim1998}, but highlights the importance of cosmic rays
(and the importance of realistic background states)
for this instability.
\begin{figure*}
\centering
 \includegraphics[width=\textwidth]{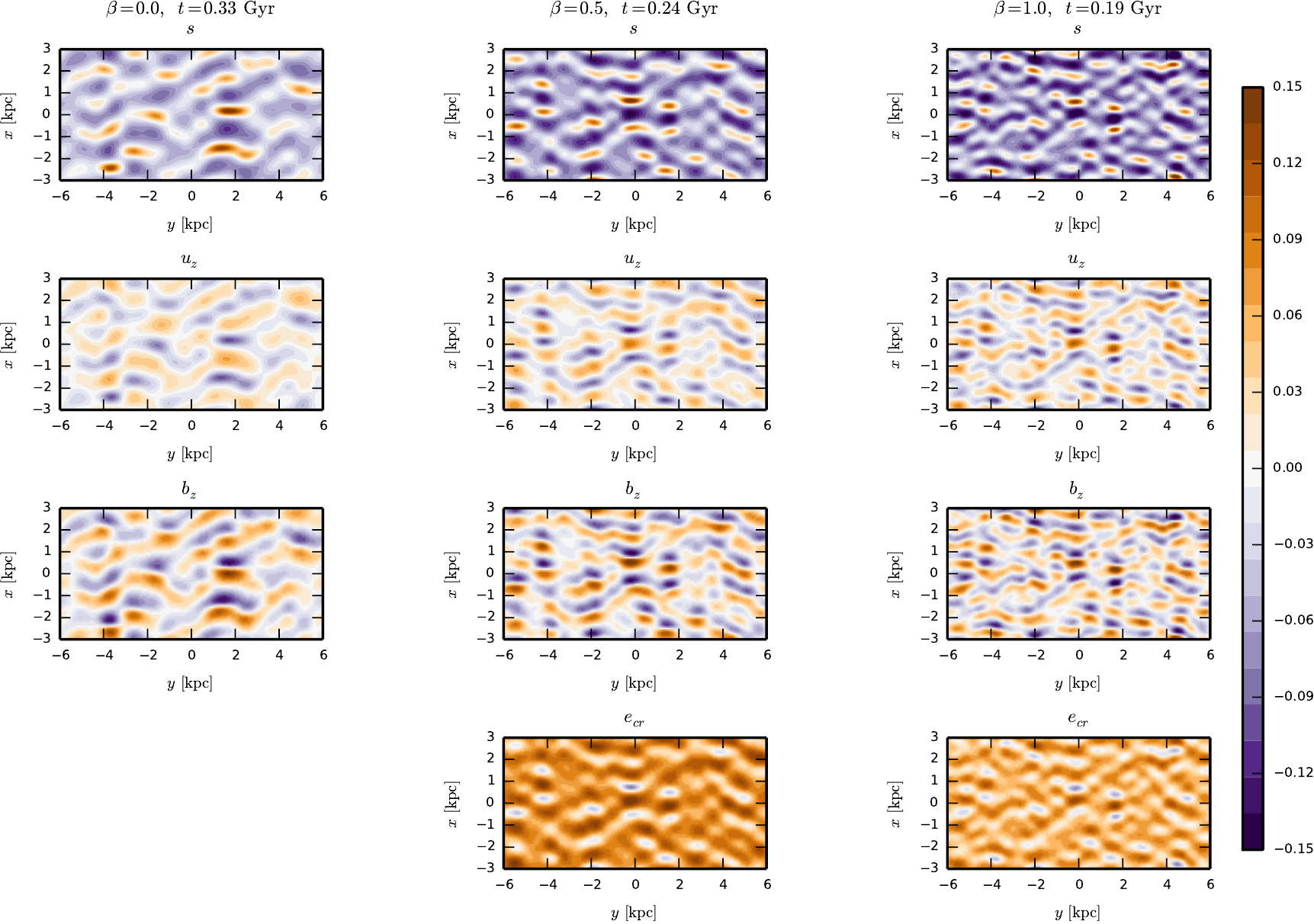}
 \caption{
             Slices at $z=0.5\kpc$ of the deviations to the initial density field,
             $s$,
             to the vertical component of the velocity field, $u_z$ and to the
             magnetic field, $b_z$ and to the cosmic rays energy density,
             $e_\text{cr}$. The snapshots at the instant when
             $E_\text{kin}=0.7 E_\text{thm}$.
         }
 \label{fig:slices_top}
\end{figure*}

Of particular interest are the disturbances to the initial state of the
system defined in Section \ref{sec:initial}. These can be expressed in terms of
dimensionless quantities, defined as
\begin{align}\label{eq:perturbations}
    s(z) =&\, \,\frac{\rho(z)-\rho_0(z)}{\rho_0(z)} \,,\\
    \bm{u}(z) =& \,\,\frac{\bm{U}(z)}{c_s}  \,,\\
    \bm{b}(z) =&\, \,\frac{\bm{B}(z)-\bm{B}_{0}(z)}{B_{0,y}(z)}   \,,
\\
    e\CR(z) =&\,
\,\frac{\ecr(z)-\epsilon_{\mathrm{cr},0}(z)}{\epsilon_{\mathrm{cr},0}(z)} \,,
    \label{eq:perturbations2}
\end{align}
where the subscript $0$ denotes the quantity at $t=0$. These have the same form as
the perturbations defined by \citet{Giz1993} and provide an estimate for the degree
of non-linearity of the system.

In Figure \ref{fig:slices_top} slices at constant
height ($z=0.5\kpc$, i.e.\ one scale-height away from the midplane) 
are shown for different choices of $\beta$.
The selected snapshots correspond to the instant when the kinetic energy 
in the domain reaches $70$ per cent of thermal energy,
to allow comparisons of the spatial structure on an equal footing
(i.e.\ to account for the different timescales of three runs), 
with the actual time indicated at the top of each column.
The system is still in the linear phase (with amplitudes typically being below
20\%) but a significant amount of non-trivial structure is already visible.
The spatial distribution of the substructures does not change with time\footnote{For
a movie, see \href{http://doi.org/8j3} {http://doi.org/8j3} .}, only their amplitudes vary.
The observed structures are comprised of over-dense sheets of inflowing gas
surrounded by outflowing, magnetized, cosmic ray-rich, under-dense gas. 

\begin{figure*}
\centering
 \includegraphics{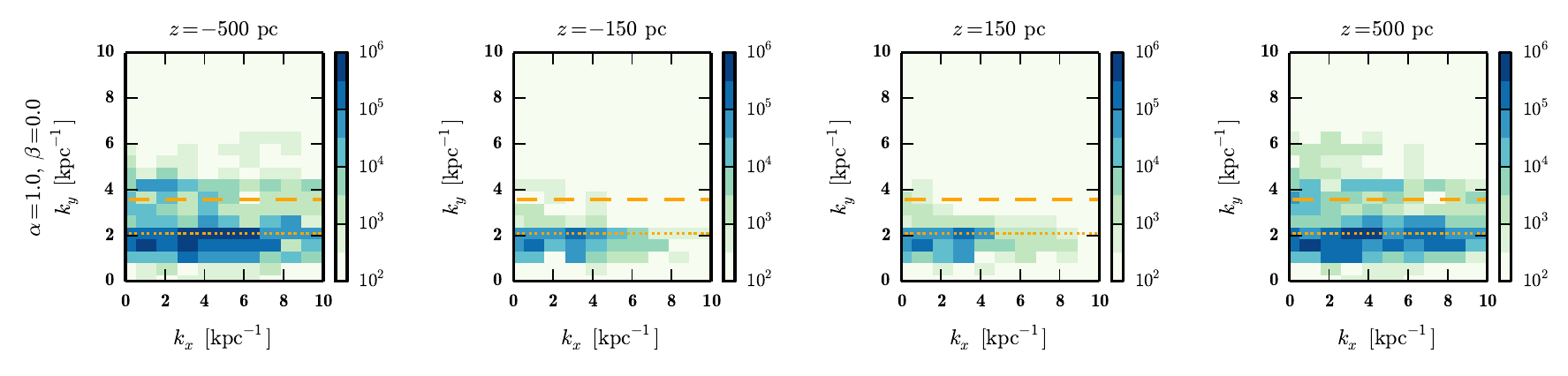}
 \includegraphics{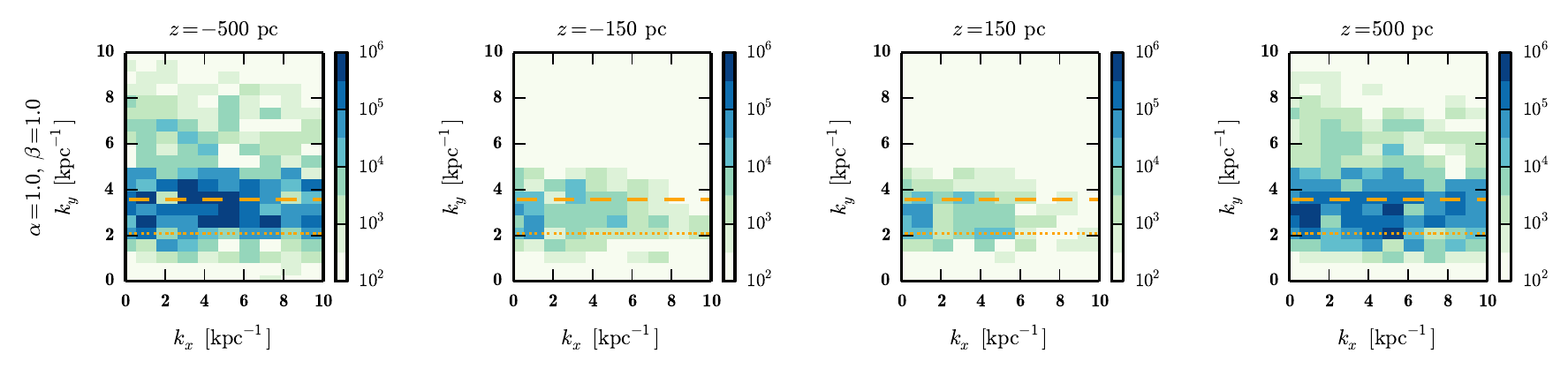}
 \caption{
             Two-dimensional power spectrum of $b_z=B_z/B_{y,0}$.
             The power spectra were computed over slices of fixed $z$, as
             indicated in each panel.
             The top row shows the power spectra for $\alpha=1$ and $\beta=0$.
             The bottom row shows the power spectra for $\alpha=1$ and
             $\beta=1$.
             Different colors correspond to different powers, as shown in the
             colorbar.
             Contour lines (drawn without taking the log) highlight major the
             peaks.
             The short- and long-dashed \revision{orange} curves show the $k_y$
             associated with
             the fastest growing mode obtained from the calculations of
             \cite{Giz1993} for $\alpha=1$ and $\beta=0,1$, respectively.
         }
 \label{fig:Pk_B}
\end{figure*}
\begin{figure*}
\centering
 \includegraphics{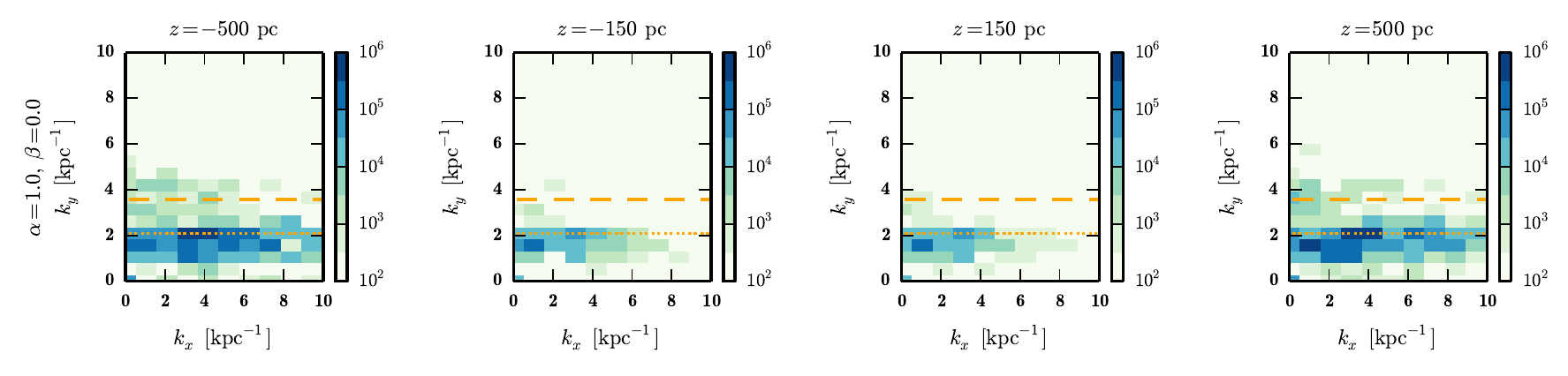}
 \includegraphics{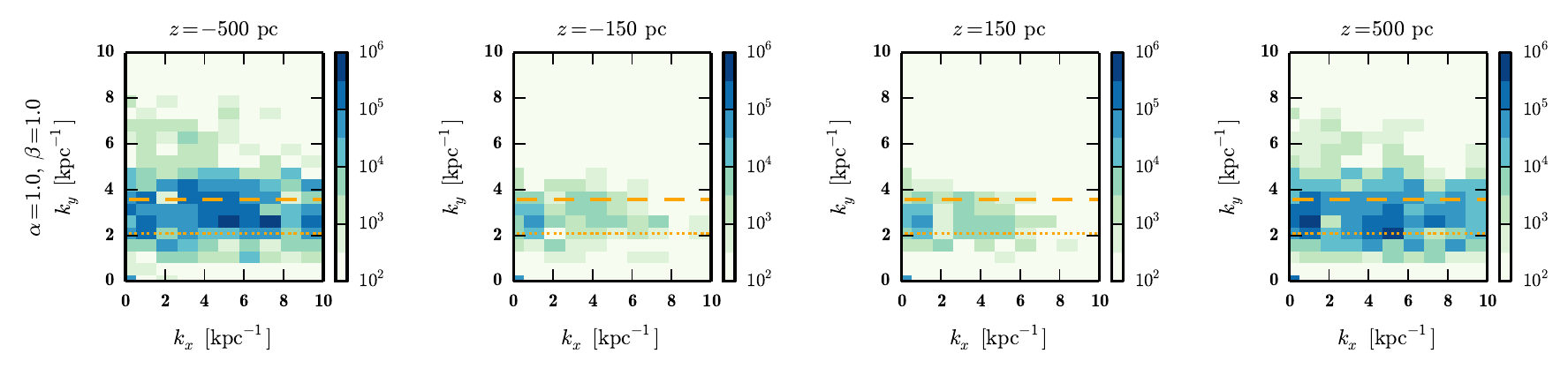}
 \caption{
             Same as Figure \ref{fig:Pk_B} but for the vertical component of the
             velocity perturbation, $u_z=U_z/c_s$.
         }
 \label{fig:Pk_u}
\end{figure*}

In Figures \ref{fig:Pk_B} and \ref{fig:Pk_u}, we examine the power spectra of
perturbations to the magnetic and velocity field, comparing the cases of
$(\alpha,\beta)=(1,0)$ and $(\alpha,\beta)=(1,1)$,
at the snapshot when
$E_\text{kin}=0.7\,E_\text{thm}$.
In the absence of
cosmic rays, the two dominant modes are 
$(k_x,k_y) = (1.0 {\kpc}^{-1}, 1.6 {\kpc}^{-1})$ and
$(3.1 {\kpc}^{-1}, 2.1 {\kpc}^{-1})$.
The $k_x$ value of the first mode actually corresponds 
\revision{to the domain size in the $x$-direction (i.e. $2\pi/L_x$).}
There is significant spread in $k_x$, which is consistent with the
weak dependence of the growth rate with wavenumber
noted in the analytic calculations.
There is less spread in $k_y$ however, with energy being concentrated close
to the wavenumber associated with the fastest growing mode of \cite{Giz1993},
$k_{y,\rm GS}$, which is the dotted curve plotted.
Our calculations favor slightly smaller wavenumbers (larger wavelengths),
which is understandable given the inclusion of non-zero diffusivities.

In the presence of cosmic rays, there is a general decrease of the wavelength
along the azimuthal direction. To facilitate the comparison, the two most prominent
modes in the $\beta=1$ case are
$(k_x,k_y) = (1.0 {\kpc}^{-1}, 2.1 {\kpc}^{-1})$ and
$(5.2 {\kpc}^{-1}, 2.6 {\kpc}^{-1})$.
While most of the energy is still close to the relevant azimuthal wavenumber
from \cite{Giz1993},
$k_y=k_{y,\rm GS}$ (here plotted as the dashed line),
there is more spread, showing that other modes
are also being significantly excited.
Again, the dominant modes tend to have wavenumbers slightly smaller than
the value for the ideal case.

Closer to the midplane there is less overall power
and less relative power in the fastest growing modes;
this is discussed in Section~\ref{sec:depz}, below.

\begin{figure*}\centering
 \includegraphics{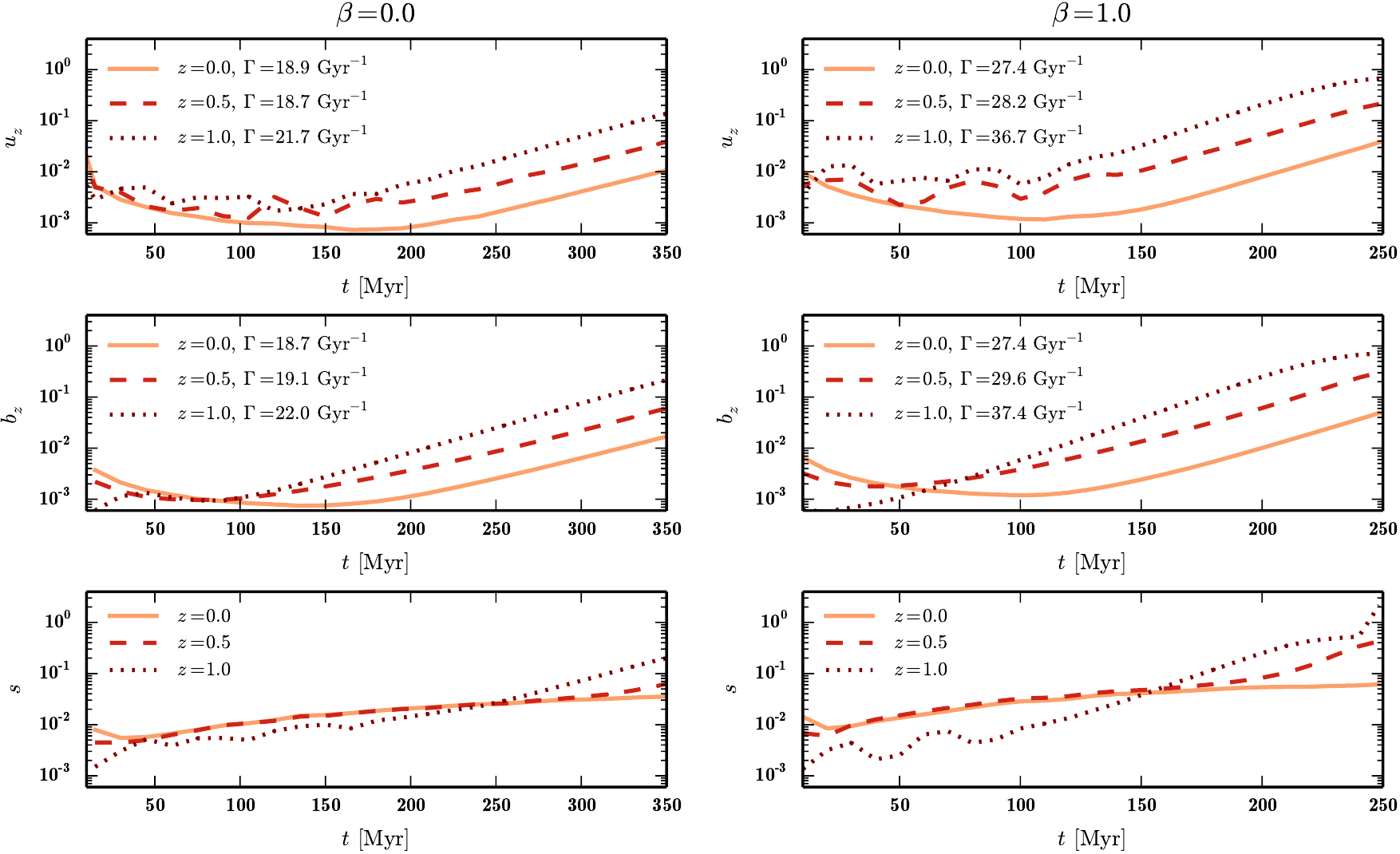}
 \caption{
          Root mean square evolution of the perturbations for models with $\alpha=1$
          at three different slices of fixed vertical coordinate, $z$, as indicated
          in each panel.
          The panels in the left column shows the results in the absence of cosmic
          rays, while the right-hand column shows the case of $\beta=1$.
         }
 \label{fig:rms_evolution}
\end{figure*}

\subsection{Dependence with height ($z$)}
\label{sec:depz}
The evolution of the rms of some components of the quantities defined in
equations \eqref{eq:perturbations}--\eqref{eq:perturbations2}
is shown in Figure~\ref{fig:rms_evolution}, for different $z$.
The vertical components of the velocity and magnetic fields
show clear exponential growth phases.
There is small, but non-negligible, variation in the growth rates calculated for
varying $z$ (with growth rates slightly smaller for low $z$).
The reason for this variation is that, as discussed in the previous Section,
these solutions do not correspond to a single unstable mode;
we are not here solving the linearized problem.
(And, in fact, even if we had a single mode, our background state is slowly
evolving away from the initial state.)
Nevertheless, these variations do not prevent a meaningful overall growth rate
for the instability being calculated.
The growth is clearest in the components $u_z$ and $b_z$ shown,
as these components are most directly driven by the instability.
Other variables (including $s$, also shown) are more subject to nonlinear effects,
including the mixing of different modes and the evolution of the
background state. The latter is a significant effect, as the average
density in the midplane ($z=0$) at the time 
$E_\text{kin} = 0.7 E_\text{thm}$ is
on average $3\%/5\%/5\%$ higher than the initial state
for the cases $\alpha=1$ and $\beta=0/0.5/1$,
due to the net sinking of gas produced
by the Parker instability. Similarly, there is a net rise of cosmic rays, with the cosmic ray 
energy density at that time being $15\%/11\%$ smaller than the initial state, for $\alpha=1$ and $\beta=0.5/1$.

In many of the linearized analyses \citep[e.g.][]{Giz1993,Kim1998},
it is noted that the instabilities may possess a specific symmetry
with respect to reflection in the midplane:
either symmetric (the so called `odd' modes, with no midplane crossings),
with $s$, $u_x$, $u_y$, $b_x$, $b_y$ and $e\CR$ symmetric 
and $u_z$ and $b_z$ antisymmetric 
under the mapping $z \rightarrow -z$;
or antisymmetric (the `even' modes, allowing midplane crossings),
with $s$, $u_x$, $u_y$, $b_x$, $b_y$ and $e\CR$ antisymmetric
and $u_z$ and $b_z$ symmetric under this mapping.
Note that these perturbations are with respect to the initial state,
which in these terms is itself symmetric (`odd').
For some systems, \citep[e.g.][]{Horiuchi1988,Basu1997}
it is found that the \revision{antisymmetric} mode of instability is preferred;
and it is notable that this allows an effective factor of two decrease
in the horizontal separation of high density patches
(since the antisymmetric density perturbation allows adjacent peaks
in the $(y,z)$-plane to be at $(y,z)$ and $(y+\lambda_y/2,-z)$;
rather than at $(y,\pm z)$ and $(y+\lambda_y,\pm z)$,
as for the symmetric case).
This has implications for the establishment of realistic spacings
in galactic density concentrations (i.e.\ giant molecular clouds);
and also for the timescales of evolution of these features,
since the linear growth rates depend upon the wavelength.
Some authors \citep[e.g.][]{Basu1997,Mouschovias2009}
consider the nonlinear evolution of the most unstable pure symmetry mode.

However, \cite{Giz1993}, \cite{Kim1998} note that
for a system with a `realistic' gravity profile,
such as we study,
the growth rates of the most unstable modes show
no preference for either type of symmetry.
And in any event, as the solution evolves nonlinearly
(and with interaction with the symmetric initial state),
the antisymmetric modes excite symmetric components,
and so purely antisymmetric perturbations cannot be sustained
in the nonlinear regime.

\revision{A simple visual inspection suggests that our} 
nonlinear solutions do not exhibit either pure
symmetry or antisymmetry
with respect to reflection in the midplane.
Note that our weak random initial velocity is asymmetric,
so does not seed either pure symmetry of perturbation.
Nevertheless, if one symmetry was genuinely preferred,
we would expect the weak initial component of the other symmetry
simply to decay.
(And were a pure symmetry solution to be obtained from a pure symmetry
initial condition, it could not be considered robust to the inevitable
fluctuations of a real system.)

\begin{figure}
 \includegraphics[width=\columnwidth]
 {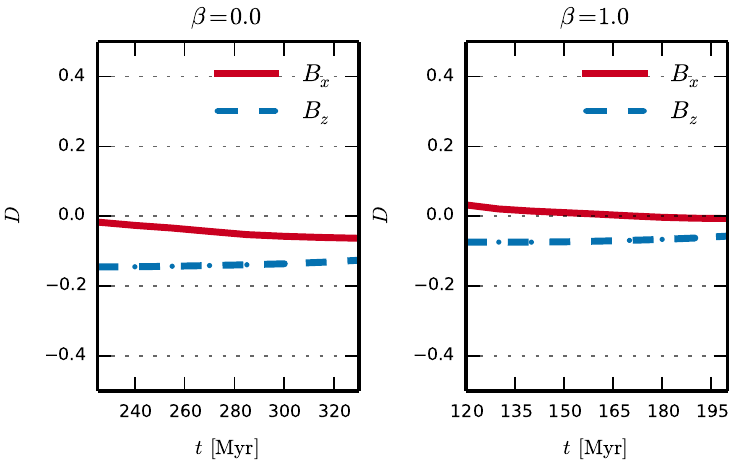}
 \caption{\revision{Time evolution of $D$ (equation \ref{eq:parity}), in the cases of
 $\beta=0$ and $\beta=1$ (with $\alpha=1$ in both cases). $D=1$ indicates pure
symmetry with respect to reflection in the midplane, while $D=-1$ indicates pure
antisymmetry.}}
 \label{fig:symmetry}
\end{figure}

\revision{To better examine this, we studied how the magnetic energy is distributed
between symmetric and antisymmetric modes. For simplicity, we restrict
the analysis to the $x$ and $z$ components, 
which are zero at the start of the simulation
and avoid the complication of the symmetry of the background state.
The symmetric part is}
\begin{equation}
 B_x^{({\rm s})}(x,y,z) = \frac{1}{2}\left(\, B_x(x,y,z) + B_x(x,y,-z) \,\right) \,, 
\end{equation}
\begin{equation}
 B_z^{({\rm s})}(x,y,z) = \frac{1}{2}\left(\, B_z(x,y,z) - B_z(x,y,-z) \,\right) \,,
\end{equation}
\revision{and the antisymmetric,}

\begin{equation}
 B_x^{({\rm a})}(x,y,z) =\frac{1}{2}\left(\, B_x(x,y,z) - B_x(x,y,-z) \,\right) \,, 
 \end{equation}
 \begin{equation}
 B_z^{({\rm a})}(x,y,z) =\frac{1}{2} \left(\, B_z(x,y,z) + B_z(x,y,-z) \,\right) \,.
\end{equation}

\revision{From these, we compute the total energy associated with each
magnetic field component,
$E_{x/z}^{({\rm s}/{\rm a})}=\int (B_{x/z}^{({\rm s}/{\rm a})})^2/8\pi\,\dv V$, in the simulation domain
and construct the following diagnostic quantity}
\begin{equation}
\label{eq:parity}
 D = \frac{E_{x/z}^{({\rm s})} - E_{x/z}^{({\rm a})}}{
	     E_{x/z}^{({\rm s})} + E_{x/z}^{({\rm a})}}\,.
\end{equation}
\revision{Thus, $D=1$ for pure symmetry or $D=-1$ for pure antisymmetry.}

\revision{In figure \ref{fig:symmetry}, the evolution of $D$ is shown. In the
$\beta = 0$ case, there is a slight preference for antisymmetric modes 
($\,0~\!\!>~\!\!D>~\!\!\!-0.15\,$). 
The presence of cosmic rays makes any preference even less clear.
}

We therefore conclude that pure symmetry with respect to reflections
in the midplane should not be expected in the nonlinear case.
In this respect, the situation is rather like that
with the horizontal structures, noted above;
in the nonlinear regime, a spectrum of modes,
of differing wavenumbers $(k_x,k_y)$ and of differing midplane-symmetries,
must be expected.
We note that this was appreciated by \cite{Giz1993},
in their discussion of continuum modes
(to which our solutions \revision{should} correspond;
\revision{since,}
as noted by \cite{Kim1998},
\revision{these} have growth rates significantly higher than the discrete modes,
so must be expected to dominate).
They observed that the analysis of individual modes is of limited relevance,
with an initial value treatment (such as that adopted here)
being more appropriate;
and with the finally realised state
depending on initial conditions and nonlinear effects.
As \cite{Giz1993} also note, however, the modal analysis remains valid
with respect to conclusions about exponential growth
(as confirmed by Figure~\ref{fig:rms_evolution}),
since we are still working with a complete set of functions.

\subsection{Dependence on diffusivities ($\nu$, $\eta$, $\kappa_\parallel$)}
\label{sec:Pm}

\begin{figure}
\centering
 \includegraphics[width=\columnwidth]{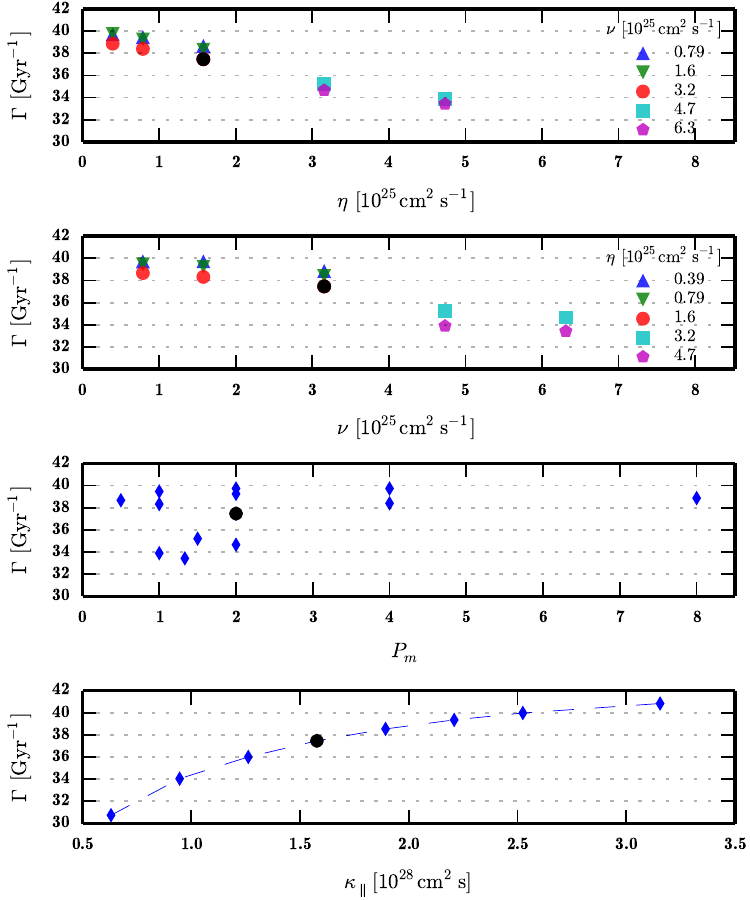}
 \caption{
            The dependence of the growth rate of $U_{rms}$ in the model with
            $\alpha=\beta=1$ on: the magnetic diffusivity, $\eta$; the
            viscosity, $\nu$; the magnetic Prandtl number, $P_m$, and the 
            cosmic ray diffusivity, $\kappa_\parallel$.
            The fiducial values are shown as black circles.
         }
 \label{fig:varyDiffs}
\end{figure}

The previous results were for our fiducial choices
of constant kinematic viscosity
$\nu_\text{fiducial}\approx 3.16 \times 10^{25}\cm^2\s^{-1}$
and constant magnetic diffusivity
$\eta_\text{fiducial}\approx 1.58 \times 10^{25}\cm^2\s^{-1}$,
corresponding to magnetic Prandtl number $\text{Pm}=\nu/\eta=2$.
The dependence of the results on these parameters was examined;
in Figure~\ref{fig:varyDiffs} we compare growth rates
with variations in the ranges $\,0.25$--$2\,\nu_\text{fiducial}$ and
$\,0.2$--$3\,\eta_\text{fiducial}$.
There are small increases in the growth rate,
$\sim\!1$--$2\Gyr^{-1}$ (or $\lesssim 5\%$),
when the magnetic diffusivity is decreased to $1/5$ or the
viscosity is reduced to $1/4$ of the fiducial values.
Conversely, increasing $\nu$ and $\eta$ by factors $2$ and $3$,
respectively, leads to decreases in the growth rate $\lesssim 10\%$.
Such a decrease of the growth rate with increasing diffusivities
might be expected on energetic grounds;
but the choice of diffusivities clearly has only a small effect
on the growth rate, particularly when compared to its variation
with $\alpha$ and (particularly) $\beta$.

Together, the variations with $\nu$ and $\eta$ show no clear dependence
on the magnetic Prandtl number in the range $\,\text{Pm}=0.5$--$8$;
as can be seen in the third panel of Figure~\ref{fig:varyDiffs}.
This range of course remains far from the expected values for galactic disks,
$\text{Pm}\gg 1$;
but along with the agreement with the ideal linearized studies,
this nevertheless suggests that the instability is not strongly
influenced by diffusion.

In the lowest panel of Figure~\ref{fig:varyDiffs} we show the impact on the 
growth rate of varying the cosmic ray diffusivity along the magnetic field,
$\kappa_\parallel$. 
In agreement with the results of \citet{Ryu2003}, 
the difference between realistic values and even larger values
(i.e.\ moving towards the approximation of infinite parallel diffusivity) 
is small:
doubling the value of $\kappa_\parallel$ from our fiducial value
leads to a $9\%$ increase in the growth rate. 
(Decreasing $\kappa_\parallel$ from the fiducial value 
gives a slightly larger effect;  
but this direction is moving away from galactically reasonable values.)

\subsection{Observational signatures}
\label{sec:obs}
\begin{figure}
 \includegraphics[width=\columnwidth]{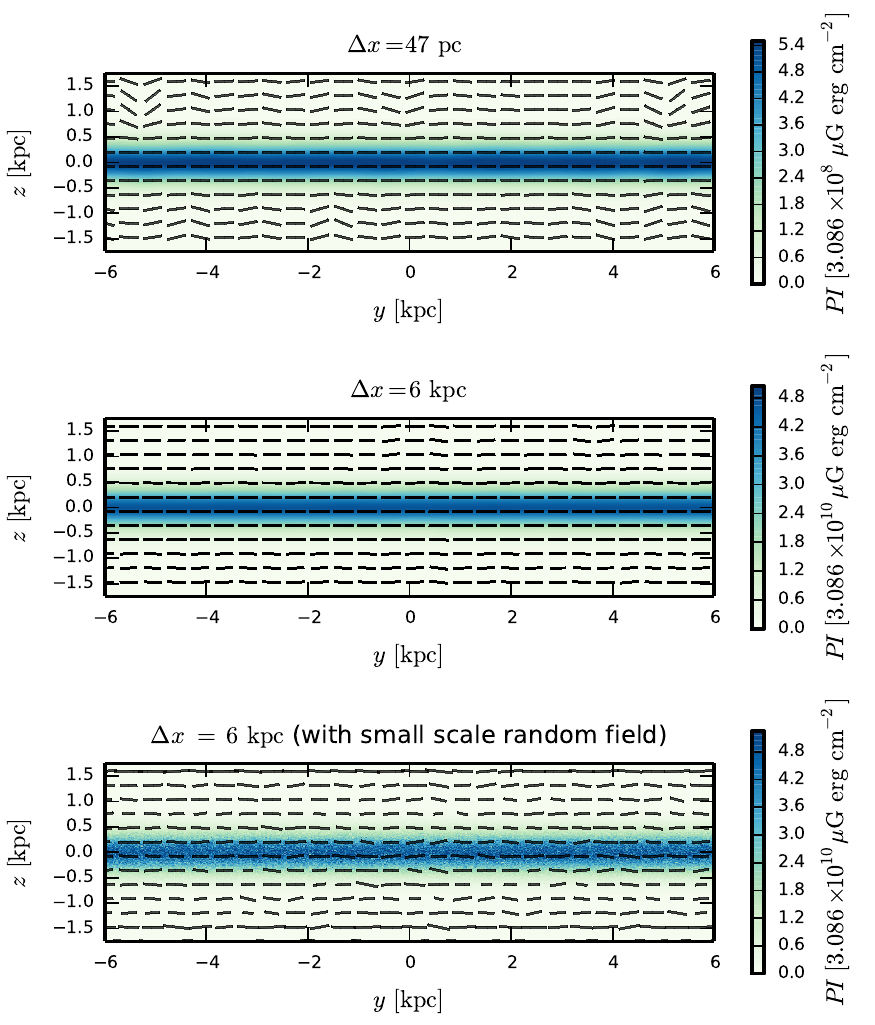}
 \caption{Polarization view of the model with $\alpha=\beta=1$. The shaded contours in the three
          panels show the polarized intensity, $PI$, in arbitrary units; dashes are
          oriented perpendicularly to the angle of polarization and their lengths
          show the degree of polarization. For each panel, the integration was
          over the interval indicated above that plot. The bottom panel show the effect
          of the inclusion of a gaussian random field.}
 \label{fig:polarization}
\end{figure}

\begin{figure}
 \includegraphics[width=\columnwidth]{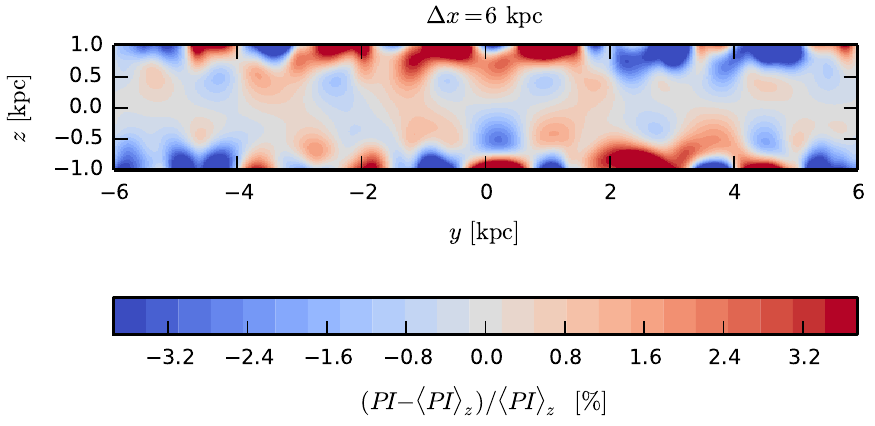}
 \caption{Fluctuations in the polarized intensity relative to the average at each 
          $z$ are shown, 
          for the solution analyzed in Figure~\ref{fig:polarization}.}
 \label{fig:polarization_differences}
\end{figure}

Having a satisfactory model where Parker instability is the dominant effect,
we now look for simple possible observational signatures and their dependence on the
cosmic ray content. We start by examining the archetypal signature of the Parker instability: the so-called Parker
arches/loops. Naively, one would expect the loops to be present in (polarized) synchrotron
intensity maps of edge-on galaxies. 
Thus, we show in Figure~\ref{fig:polarization} 
polarized synchrotron intensity at $\lambda=1\cm$ 
(see appendix \ref{ap:polarization} for the details of the calculation). 
In the top panel, which shows the result of
integrating over a thin slice of width $\Delta x=47\pc$,
the polarization angles are affected by the undulations 
of the magnetic field lines. 
The polarized and total synchrotron intensities, however, 
are dominated by the disk stratification, 
and the variations at the same disk height are almost negligible.
Observing an isolated thin slice of an edge-on galactic disk 
is not possible. Thus, 
in the second panel the integration is performed over the whole simulation box
($\Delta x = 6\kpc$). The undulations in the polarization angles are mostly 
erased and the signal in the intensity remains almost featureless. 
In Figure~\ref{fig:polarization_differences} 
the fractional difference to the average at each $z$ is shown: 
a very weak signal ($\lesssim2\%$) can be identified within one 
scale height from the midplane; 
at larger distances stronger fluctuations relative to 
the $z$-average are found ($\sim3.5\%$).
Finally, in the third panel of Figure \ref{fig:polarization}, 
we check the effect of the presence of a turbulent random field component,
by adding to each point a gaussian random field with a root
mean square value equal to the magnitude of the model field at that height $z$.
The presence of the random field removes any remaining (weak) signal of the Parker instability in the polarization.

The prospect of obtaining direct evidence of Parker loops from this kind of 
synchrotron
observations is further complicated by the curvature of the galaxy, which would
further weaken the signal after the integration.
Figure \ref{fig:polarization} also assumes the observations are made at a
wavelength of $1\cm$, which minimizes Faraday rotation effects.

\begin{figure*}
\centering
 \includegraphics{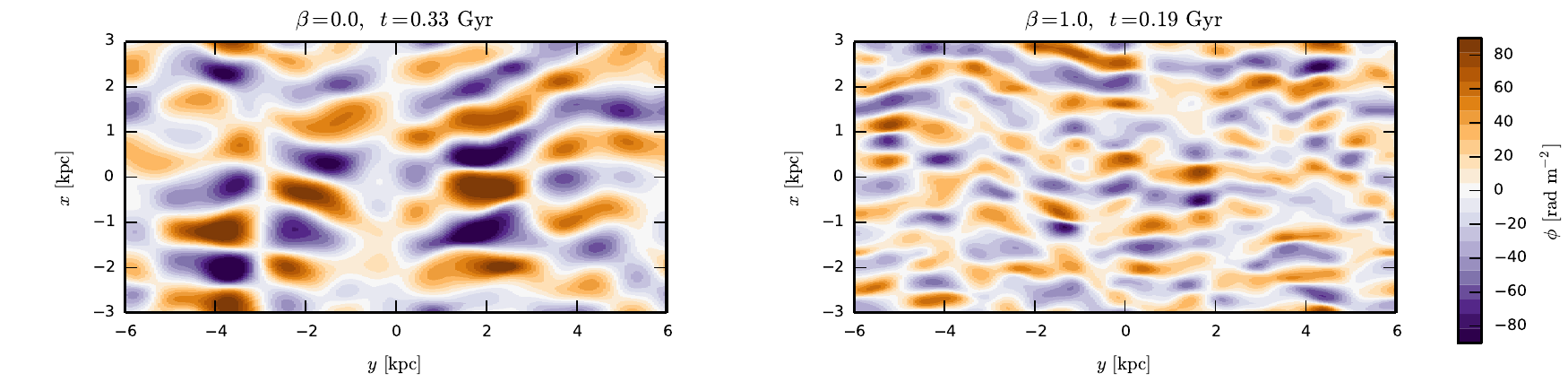}
 \caption{
           Faraday rotation measure maps for different $\beta$ as indicated 
           (for $\alpha=1$), at corresponding points in the evolution of the instability
           (when $E_\text{kin}=0.7 E_\text{thm}$).
         }
 \label{fig:RM}
\end{figure*}

An alternative approach is to consider the rotation measure (RM) to probe 
the fluctuations in the density and magnetic field produced by the Parker instability (Figure \ref{fig:slices_top}). 
Assuming the case of a face-on galaxy, the RM can be computed from
\begin{equation}
 \phi  = (0.812\,\rad\m^{-2}) \int \frac{n_e(z)}{1\cm^{-3}} \frac{B_z}{1\muG}
\,\frac{\dv z}{1 \pc}\,, 
\end{equation}
where $n_e$ is the number density of electrons in the path (for simplicity, we 
assumed the plasma was completely ionized).

\begin{figure}
\centering
 \includegraphics[width=0.98\columnwidth]{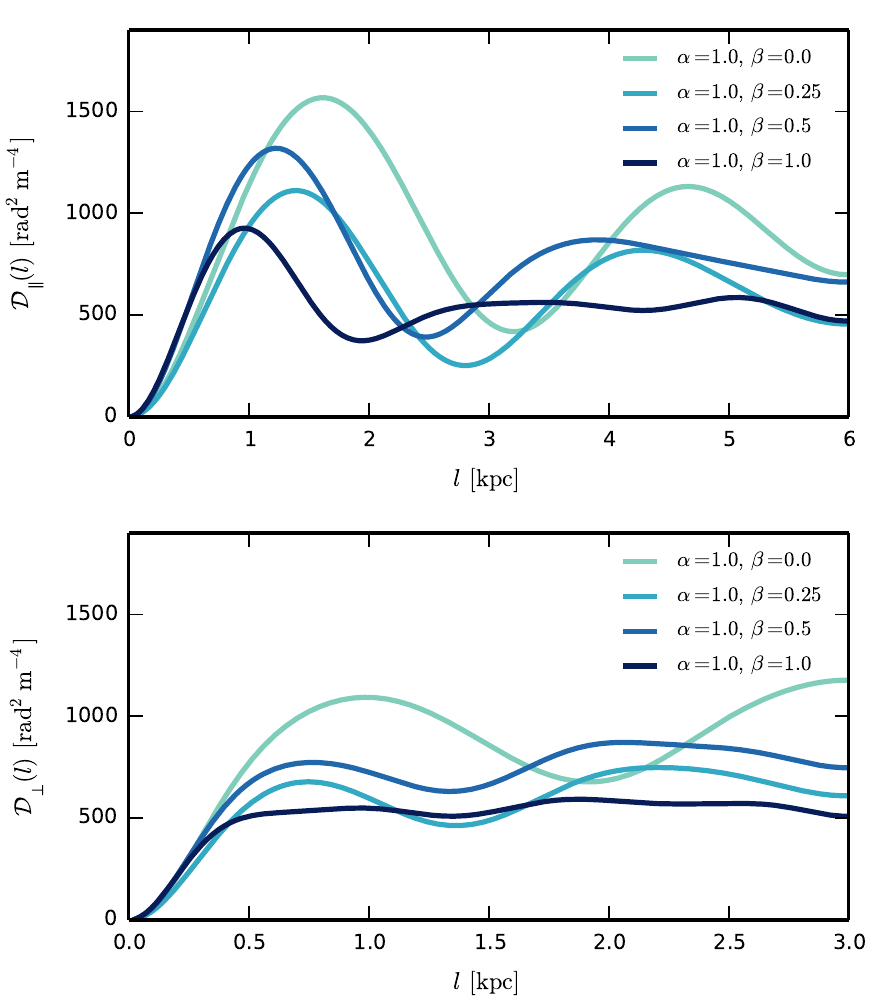}
 \caption{
           Structure functions computed from Faraday rotation measure maps
           for runs with $\alpha=1$ and $\beta$ varying 
           as indicated in the legend.
           The top panel shows the structure function along the $x$-axis
           (i.e.\ perpendicular to the initial magnetic field)
           while the bottom panel shows the structure function along the
           $y$-direction.
         }
 \label{fig:RM_sf}
\end{figure}

In Figure \ref{fig:RM}, Faraday maps obtained from the simulation 
are shown. The integration was
performed over the whole domain (i.e.\ a distant external radio source is
assumed). There are sharp peaks in the rotation measure signal, with a maximum
$|\phi|\approx 85\,\rad^2\m^{-4}$ in the $\beta=0$ case;
$|\phi|\approx 74\,\rad^2\m^{-4}$ in the $\beta=0.5$ case; and
$|\phi|\approx 91\,\rad^2\m^{-4}$ in the $\beta=1$ case.
One should, however bear in mind that these amplitudes correspond to a
particular choice of output time (namely, the time when
$E_\text{kin}=0.7\,E_\text{thm}$) and that with the current setup no steady
state is reached.
More informative is the spatial distribution of the RM structures,
which changes negligibly with time.
In Figure \ref{fig:RM_sf} the structure function (SF) of the RM is
plotted, computed parallel and perpendicular to the direction of the initial
magnetic field ($y$-direction), i.e.\
\begin{align}
\mathcal{D}_\parallel(l) =&
    \langle \left[ \phi(x,y)-\phi(x,y+l) \right]^2 \rangle\,,\\
\mathcal{D}_\perp(l) =&
    \langle \left[ \phi(x,y)-\phi(x+l,y) \right]^2 \rangle\,.
\end{align}

There are systematic differences in the SFs along and perpendicular to the large
scale field, and features which are dependent on $\beta$. The first peak of the
SF is generally larger in the direction along the field than in the direction
perpendicular to it: for $\beta=0$, the first peak occurs at
$l_\text{peak}\approx\!1\kpc$ for $\mathcal{D}_\perp$,
while $l_\text{peak}\approx\!1.5\kpc$ for  $\mathcal{D}_\parallel$. When $\beta$ is
increased, the peaks of the SF parallel to the field are shifted towards
smaller scales, with $l_\text{peak}\approx0.95\kpc$ for $\beta=1$. In the
direction perpendicular to the initial field, the SF is flattened with the
increase in the cosmic ray density.

\begin{figure}
\centering
 \includegraphics[width=0.98\columnwidth]{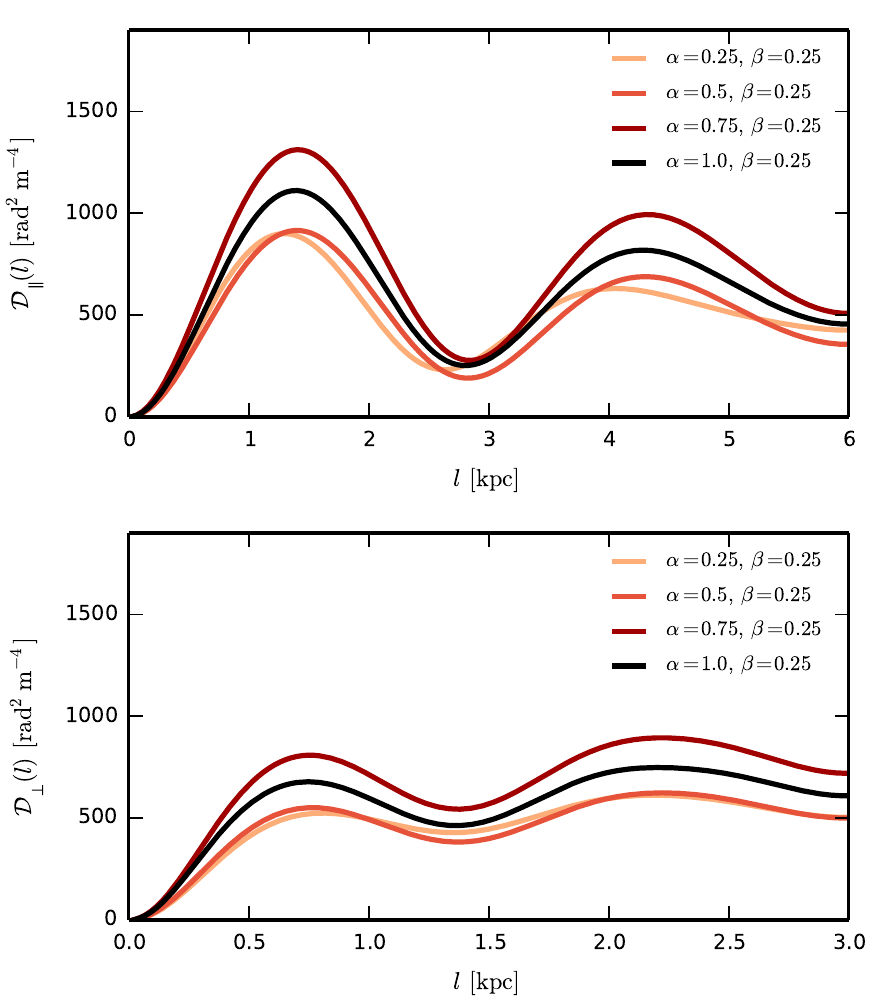}
 \caption{
           Structure functions computed from Faraday rotation measure maps
           for runs with $\beta=0.25$ and $\alpha$ varying
           as indicated in the legend.
           Top panel shows the structure function along the $x$-axis
           (i.e.\ perpendicular to the initial magnetic field)
           while the bottom panel shows the structure function along the
           $y$-direction.
         }
 \label{fig:RM_sf_alpha}
\end{figure}

In Figure \ref{fig:RM_sf_alpha} this is examined for varying $\alpha$.
There is negligible variation on the peak positions if only the
magnetic field strength is varied.
Although this may be surprising,
it is consistent with the relatively weak effect
of varying magnetic buoyancy shown in Figure~\ref{fig:growth},
and again highlights the importance of incorporating cosmic rays
into these studies.


\section{Conclusions}
\label{sec:conc}

We have performed a set of three-dimensional numerical simulations
exploring the growth rates and length scales
associated with the Parker instability,
in a section of a galactic disk with realistic vertical structures,
and including cosmic rays via a fluid approximation.
Aiming for clarity and to avoid any bias (via unintended forcing),
our system is allowed to evolve passively from its initial state;
i.e.\ there is no artificial maintenance of
cosmic rays and magnetic fields by imposed sources.
This evolution of the background state represents a slight difference
from the linearized systems investigated in related analytical studies,
but the evolution is very slow compared with the timescales of the Parker
instability;
and we differ from such studies in more fundamental ways, in any event,
incorporating finite diffusivities and retaining nonlinear interactions.

Nevertheless, the growth rates and length scales we obtain ---
and their dependences on the model parameters ---
qualitatively agree well with such linearized studies.
The quantitative differences can be clearly understood in terms of
the differences between the models (the linearized studies being ideal,
and in some cases using alternative background profiles),
or simply due to the nonlinear nature of our solutions.

Our calculations invariably produce non-trivial spatial structures,
significantly involving multiple modes in the instability,
and with these modes being fully three-dimensional 
(rather than pure undular or interchange modes)
\revision{and with no preference for either pure symmetry about the midplane}.
This underlines the importance of three-dimensional calculations,
if signatures of the instability structures,
for comparison with observations, are sought.
And the clear effect of the cosmic rays on the growth rate and wavelengths
of the instability (as seen when varying our parameter $\beta$),
highlights the importance of including this ingredient of the ISM
in such calculations.

Although our calculations use high values
for viscous and magnetic diffusivities,
in comparison with galactic disks,
our investigation with varying values suggests that the instability
is not being significantly affected by these diffusivities.

For simplicity, but in contrast to real galaxies,
our calculations do not involve rotation.
Rotation is normally found to stabilize against the Parker instability,
but we note that \cite{Kim1998} argue that this will be less the case for
a cosmic ray-driven instability,
cf.\  instabilities with purely magnetic and gas buoyancy.
And preliminary runs with rotation (not reported here)
confirm that the effect is minor.
Similarly, our calculations do not involve rotational shear
(differential rotation), which is significant in disk galaxies.
\citet{FT94,FT95} find differential rotation to weaken the Parker instability;
the specific effects on our calculations should be considered in future work.

As a first step towards investigating the observational implications,
we calculate synthetic polarized intensity and Faraday rotation measures
maps from our simulations, and compute the structure functions associated with
the latter.
We find that it is very unlikely that radio observations of edge-on galaxies 
would be able to clearly detect any structures produced by the instability,
because of the averaging along the line-of-sight and the masking of any
signal by the presence of the random (small-scale) magnetic field.
However, our results suggest there may be strong signatures in
Faraday rotation measures of face-on galaxies. 
And our preliminary analysis suggests that,
when combined with independent data about the disk scale height,
the correlation scales inferred from such rotation measure maps
may be a useful observational diagnostic
for the cosmic ray content of galaxies.

A comprehensive study based on the present model, solely dedicated to the 
observational signatures of Parker instability in galaxies, is underway.

\newpage
\section*{Acknowledgements}
\revision{We thank the referee for many constructive comments on the original 
version of this paper.}
We also thank Ann Mao for useful discussions.
LFSR has been supported by STFC (grant ST/L005549/1) and acknowledges support
from the European Commission's Framework Programme 7, through the Marie Curie
International Research Staff Exchange Scheme LACEGAL (PIRSES-GA-2010-269264).
\revision{This work was partially supported by the Leverhulme Trust 
(PRG-2014-427).}
This work made use of the facilities of N8 HPC provided and funded by the
N8 consortium and EPSRC (Grant No.EP/K000225/1). The Centre is co-ordinated
by the Universities of Leeds and Manchester.
This research has made use of NASA's Astrophysics Data System.

\appendix
\section{Comparison with a larger domain}
\label{ap:tall_box}
\begin{figure}
\centering

\includegraphics[width=0.495\columnwidth]
{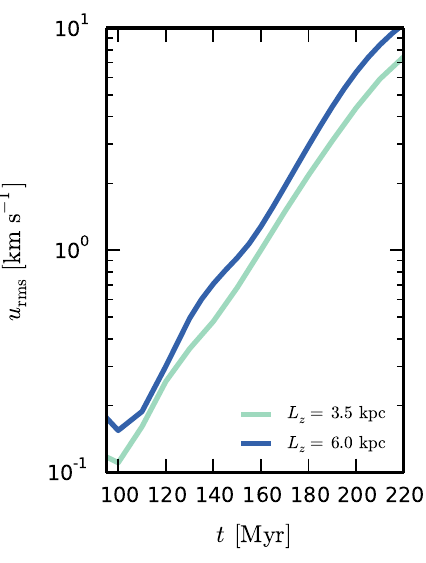}
\includegraphics[width=0.495\columnwidth]
{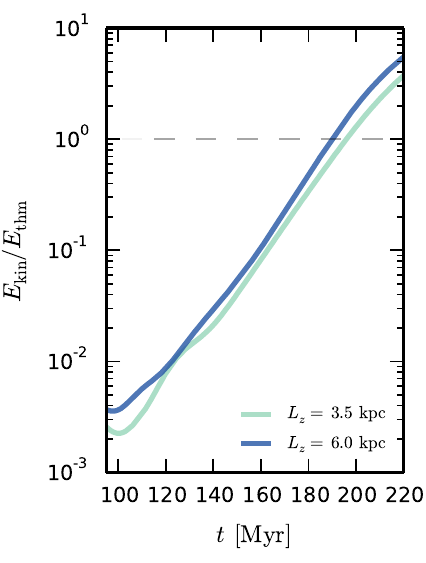}
 \caption{\revision{
             Evolution of the root mean square velocity (left panel) and
             kinetic energy (right panel), comparing the standard
             simulation domain, $(L_x,\,L_y,\,L_z)=(6\kpc,\,12\kpc,\,3.5\kpc$), with
             an enlarged domain $(L_x,Ly,L_z)~\!=~\!(6\kpc,12\kpc,6\kpc)$, for
             $\alpha=\beta=1$.
         }}
 \label{fig:growth_tall}
\end{figure}
\begin{figure*}
\centering
 \includegraphics{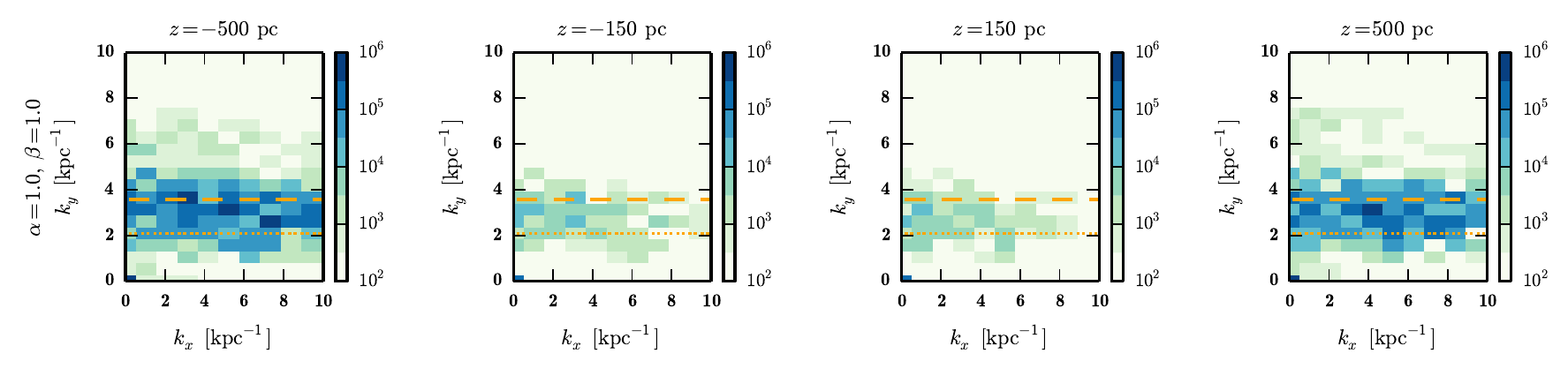}
 \includegraphics{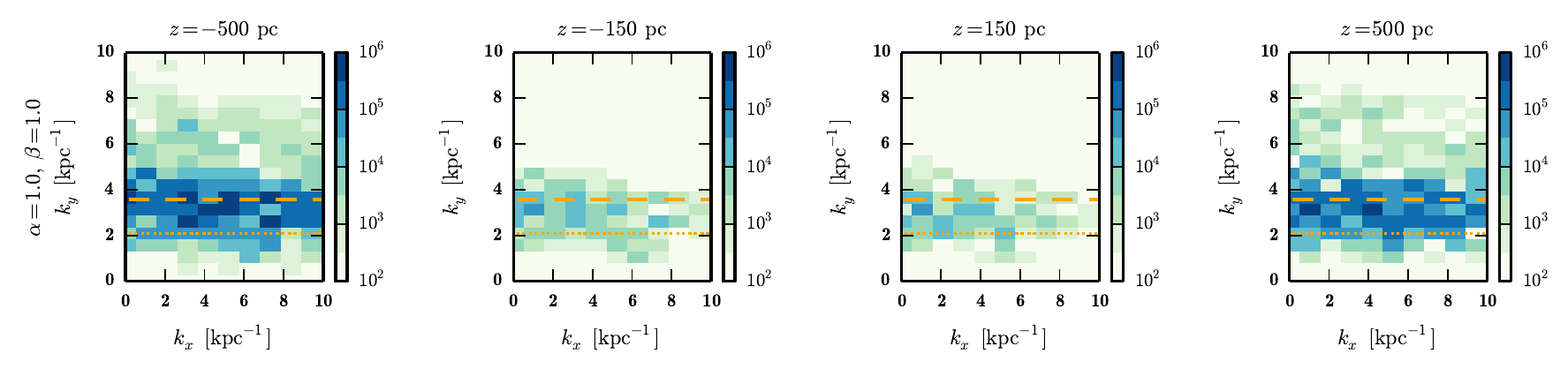}
 \caption{\revision{
             Two-dimensional power spectra computed for a simulation with an
             enlarged domain in the z-direction ($L_z=6\kpc$), for
             $\alpha=\beta=1$.
             The top row shows the power spectrum of $u_z=U_z/c_s$ while
             the bottom row shows the power spectrum $b_z=B_z/B_{y,0}$.
         }}
 \label{fig:Pk_tall}
\end{figure*}
\revision{To ensure that the vertical size of the domain is large enough to
avoid boundary effects, the simulation was also run in a larger domain,
with a vertical range $-3\kpc<z<3\kpc$, which corresponds to $6$
scale-heights between the midplane and the boundary (in contrast with the $3.5$
scale-heights in the standard runs).}

\revision{We find no significant differences between the taller
simulation domain and the fiducial one. In figure \ref{fig:growth_tall}, it can be
seen that the impact on the evolution of the kinetic energy and on the rms of the
velocity field is small. The growth rate of $U_\text{rms}$ is, however, 
$\sim\!6\%$ higher in the taller domain for the reasons discussed in section
\ref{sec:alphabeta}.}

\revision{When comparing figure \ref{fig:Pk_tall} with figures \ref{fig:Pk_B}
and \ref{fig:Pk_u}, it can be seen that the enlargement of
the vertical dimension leads only to very small changes in the power spectra;
and
these can be attributed to small variations in the random initial conditions.}

\section{Polarization properties}
\label{ap:polarization}

We discuss here the details of the calculation of the polarized intensity 
in section \ref{sec:obs}. 

First the Stokes parameters are computed,
\begin{align}
I(y,z) =& \int_0^{\Delta x} s(\bm{r}') \dv x'\,,\label{eq:I}\\
Q(y,z) =& 
\int_0^{\Delta x} p_0\, s(\bm{r}') \cos\left[2\psi(\bm{r}')\right]\dv x'\,,
\label{eq:Q}\\
U(y,z) =&
\int_0^{\Delta x} p_0\, s(\bm{r}') \sin\left[2\psi(\bm{r}')\right]\dv x'\,,
\label{eq:U}
\end{align}
where the intrinsic polarization degree was assumed to be $p_0=0.75$, 
and the local polarization angle is obtained from
\begin{align}
 \psi(\bm{r}) =& \frac{\pi}{2}+\arctan\left[\frac{B_z(\bm{r})}{B_y(\bm{r})}\right]\\
      +& 0.81\rad \left(\frac{\lambda}{1\m}\right)^2\int_x^{\Delta x}
        \left[\frac{n_e(\bm{r}')}{1\cm^{-3}}\right]
        \left[\frac{B_x(\bm{r}')}{1\muG}\right]
        \left(\frac{\dv x'}{1 \pc}\right)\,,\nonumber
\end{align}
To compute the emissivity it was assumed that the energy density of cosmic rays
is proportional to the number density of cosmic rays, thus the quantity 
\begin{equation}
s(\bm{r}) = \ecr(\bm{r}) \left[B_y^2(\bm{r})+B_z^2(\bm{r})\right]\,.
\end{equation}
corresponds to the emissivity in arbitrary units.

From equations~\eqref{eq:I}--\eqref{eq:U} it is then possible to compute the 
quantities shown in Figures \ref{fig:polarization} and 
\ref{fig:polarization_differences}. The polarized intensity is given by
\begin{equation}
 PI = \sqrt{(Q^2+U^2)}\,,
\end{equation}
the observed polarization angle is 
\begin{equation}
 \Psi = \frac{1}{2}\arctan\left(\frac{U}{Q}\right)\,,
\end{equation}
and the polarization degree
\begin{equation}
 p = PI/I\,.
\end{equation}

\bibliography{parker_instability}

\end{document}